\documentclass[preprint,showpacs,amsmath]{revtex4}

\usepackage{times}
\usepackage{hyperref}

\usepackage{graphicx}
\usepackage{dcolumn}
\usepackage{bm}

\begin{document}

\title{
Numerical precision radiative corrections to the Dalitz plot of baryon semileptonic decays including the spin-momentum
correlation of the decaying and emitted baryons
}

\author{
J.\ J.\ Torres
}

\affiliation{
Escuela Superior de C\'omputo del IPN, Apartado Postal 75-702, M\'exico, D.F. 07738, M\'exico
}
\author{
Rub\'en Flores-Mendieta
}

\affiliation{
Instituto de F{\'\i}sica, Universidad Aut\'onoma de San Luis Potos{\'\i}, \'Alvaro Obreg\'on 64, Zona Centro, San Luis
Potos{\'\i}, S.L.P.\ 78000, M\'exico
}

\author{
M.\ Neri, A.\ Mart{\'\i}nez
}
\affiliation{
Escuela Superior de F\'{\i}sica y Matem\'aticas del IPN, Apartado Postal 75-702, M\'exico, D.F.\ 07738, M\'exico
}

\author{
A.\ Garc{\'\i}a
}

\affiliation{
Departamento de F{\'\i}sica, Centro de Investigaci\'on y de Estudios Avanzados del IPN, Apartado Postal 14-740, M\'exico,
D.F.\ 07000, M\'exico
}

\date{\today}

\begin{abstract}
We calculate the radiative corrections to the angular correlation
between the polarization of the decaying and the direction of the
emitted spin one-half baryons in the semileptonic decay mode to
order $(\alpha/\pi)(q/M_1)$, where $q$ is the momentum transfer
and $M_1$ is the mass of the decaying baryon. The final results
are presented, first, with the triple integration of the
bremsstrahlung photon ready to be performed numerically and,
second, in an analytical form. A third presentation of our results
in the form of numerical arrays of coefficients to be multiplied
by the quadratic products of form factors is discussed. This
latter may be the most practical one to use in Monte Carlo
simulations. A series of crosschecks is performed. This paper is
organized to make it accessible and reliable in the analysis of
the Dalitz plot of precision experiments involving heavy quarks
and is not compromised to fixing the form factors at predetermined
values. It is assumed that the real photons are kinematically
discriminated. Otherwise, our results have a general
model-independent applicability.
\end{abstract}

\pacs{14.20.Lq, 13.30.Ce, 13.40.Ks}

\maketitle

\section{Introduction}

The radiative corrections (RC) to spin one-half baryon
semileptonic decays face three levels of complications. Despite
the important progress achieved in the understanding of the
fundamental interactions with the Standard Model \cite{part}, no
first principle calculation of these corrections is yet possible.
RC become then committed to model dependence and, what is worse,
experimental analyses which use these calculations become model
dependent, too. The second level comes from the fact that RC
depend on the process characteristics, such as charge assignment
of the baryons, type of the emitted charged lepton, size of the
momentum transfer $q$ involved, and whether real photons can be
experimentally discriminated or not. RC also depend on the
observable which is to be measured. All this requires RC to be
recalculated every time the process characteristics and the
observables are changed. At the third level one finds
complications of a practical nature. It turns out that the final
results of RC calculations are rather very inefficient to use or
are long and tedious to the point that their use becomes
unreliable. Fortunately, all the above complications can be solved
rather satisfactorily.

Although the model dependence of the virtual RC cannot be
eliminated, an extension to baryon semileptonic decays of an
analysis of Sirlin \cite{sirlin} of these corrections in neutron
beta decay shows that to orders $(\alpha/\pi)(q/M_1)^0$ and
$(\alpha/\pi)(q/M_1)$, where $M_1$ is the mass of the decaying
baryon, the corresponding model-dependence amounts to several
constants. These constants can all be absorbed into the already
present form factors of the weak decay vertex. In addition, the
theorem of Low \cite{low} in its presentation by Chew \cite{chew}
can be used to show that to these two orders of approximation the
bremsstrahlung RC depend only on the non-radiative form factors
and on the static electromagnetic multipoles of the particles
involved. Accordingly, no model-dependence is introduced in this
other part of the RC. Within these orders of approximation it is
then possible to obtain final expressions for the RC that can be
used in model-independent experimental analyses. The price is that
it will be the effective form factors (which may be indicated by
putting a prime on them) that can be experimentally determined.
The separation of the original form factors from the model
dependence of RC is then a theoretical problem only. It is in this
sense that the first level of complications is put under control.

To deal with the second level one must make an effort to calculate
RC in a way as general as possible and to be able to obtain
results which can be used directly to obtain the final results of
other possible baryon semileptonic decays. In a recent publication
\cite{rfm02} we showed that of the six allowed charge assignments
to the baryons when heavy quarks are involved, namely, $A^-
\rightarrow B^0 l^- \overline{\nu}_l$, $A^0 \rightarrow B^+l^-
\overline{\nu}$, $A^+ \rightarrow B^0 l^+ \nu_l$, $A^0 \rightarrow
B^- l^+ \nu_l$, $A^{++} \rightarrow B^+ l^+ \nu_l$, and $A^+
\rightarrow B^{++}l^- \overline{\nu}_l$, it is necessary only to
calculate the RC to the first two, to which we shall refer to as
charged decaying baryon (CDB) and neutral decaying baryon (NDB),
respectively. The RC to the other four cases are then obtained
from these two. We also showed that this property is valid up to
order $(\alpha/\pi)(q/M_1)$, when $l=e^\pm$, $\mu^\pm$, and
$\tau^\pm$ and for any observable of baryon semileptonic decays.
The problem of calculating RC is reduced considerably this way,
although it will still be necessary to recalculate for different
observables and whether real photons are discriminated or not in
the first two cases.

The third level of complications has been dealt with by computing
analytically the triple integrals over the real photon variables.
A numerical calculation of these integrals makes the application
of RC to a Monte Carlo simulation practically impossible, because
every time the values of the kinematical variables are varied
those integrals must be recalculated. The analytical form of RC
solves this problem. However, the results are very long and
tedious and the use of this latter form may become unreliable. To
control this it is very important that the analytical result be
well organized and that it be crosschecked with the triple
numerical integration form. A successful crosscheck allows the
user to gain confidence on the analytical result and on its
feeding into a Monte Carlo simulation. It may still be convenient
to find a third presentation of RC, which would make their use
more practical.

From the above discussion it is clear that the calculation of RC
to baryon semileptonic decays must be done following a program
(see Ref.~\cite{rfm02} and references therein). In previous
publications we obtained the RC to the Dalitz plot of unpolarized
decaying baryons up to order $(\alpha/\pi)(q/M_1)$
\cite{tun91,tun93}. In Ref.~\cite{rfm97} we calculated to order
$(\alpha/\pi)(q/M_1)^0$ the RC to the Dalitz plot with the angular
correlation ${\hat {\mathbf s}_1} \cdot {\hat {\mathbf p}}_2$ when
the initial baryon is polarized along ${\hat {\mathbf s}_1}$ and
the final baryon in emitted along ${\hat {\mathbf p}}_2$

In the present paper we want to attain two goals. The first one is
to continue with our program and to calculate to order
$(\alpha/\pi)(q/M_1)$ the RC to the differential angular
correlation ${\hat {\mathbf s}_1} \cdot {\hat {\mathbf p}_2}$. The
second one is to present the RC in the form of numerical arrays
which should be applied to the quadratic products of form factors
that appear in the RC, up to order $(\alpha/\pi)(q/M_1)$. We shall
cover both CDB and NDB cases.

The ordering of the paper is as follows. In Sec.~\ref{sec:virtual}
we present the results to order $(\alpha/\pi)(q/M_1)$ for the
virtual RC. In Sec.~\ref{sec:bremss} we give the results for the
bremsstrahlung RC in the triple numerical integration form and
combine them with the virtual RC results to obtain our first main
result. In addition, we give the corresponding fully analytical
results. In Sec.~\ref{sec:cks} we perform several crosschecks  and
compare with other published results. In Sec.~\ref{sec:numerical}
we proceed towards our second goal. The last Sec.~\ref{sec:diss}
is dedicated to a summary and to concluding remarks.

In order not to obscure the physics we have moved to Appendices A,
B, and C the very many algebraic expressions that appear in the
analytical results. In this paper we exhibit only new expressions.
However, previously published expressions are required in these
results. We do not reproduce them here. Instead, we give all the
necessary references so that the reader can identify them
correctly. The text and these appendices are organized so as not
to obscure the physics and to make accessible the use of our
results. Performing the analytical integrals is long an tedious.
In order to help the reader interested in checking our results we
have introduced the Appendix D, where the previous and new
integrals can be identified. In Secs.~IV and V we provide
several numerical tables with the purposes of illustration and,
more importantly, of helping the user to check his numerical
results with ours.

\section{Virtual radiative corrections}\label{sec:virtual}

Our first purpose in this section is to review our notation and
conventions. Next we shall discuss the virtual RC to the ${\hat
{\mathbf s}_1} \cdot {\hat {\mathbf p}_2}$ angular correlation
over the Dalitz plot to order $(\alpha/\pi)(q/M_1)$. The
uncorrected transition amplitude ${\mathsf M}_0$ for the baryon
semileptonic decays
\begin{equation}
A \rightarrow B \; l \; \overline{\nu}_l \label{eq:e1}
\end{equation}
is
\begin{eqnarray}
{\mathsf M}_0 = \frac{G_V}{\sqrt 2} [\overline{u}_B(p_2) W_\mu(p_1,p_2) u_A(p_1)] [\overline{u}_l(l) O_\mu v_\nu(p_\nu)].
\label{eq:e2}
\end{eqnarray}
$G_V$ is the Fermi decay constant multiplied by the appropriate
Cabibbo-Kobayashi-Maskawa factor \cite{part}. $A$ and $B$ are spin
one-half baryons, $l$ is the charged lepton, and $\nu_l$ is the
accompanying antineutrino or neutrino as the case may be. $u_A$,
$u_B$, $u_l$ or $v_l$, and $v_\nu$ or $u_\nu$ are their
corresponding spinors. The weak interaction vertex is
\begin{eqnarray}
W_\mu (p_1,p_2) & = & f_1(q^2) \gamma_\mu + f_2(q^2) \sigma_{\mu \nu} \frac{q_\nu}{M_1} + f_3(q^2) \frac{q_\mu}{M_1}
\nonumber \\
&  & \mbox{} + \left[g_1(q^2) \gamma_\mu + g_2(q^2) \sigma_{\mu \nu} \frac{q_\nu}{M_1} + g_3(q^2) \frac{q_\mu}{M_1} \right]
\gamma_5, \label{eq:e3}
\end{eqnarray}
where $O_\mu = \gamma_\mu (1+\gamma_5)$, $\sigma_{\mu\nu}=(1/2)(\gamma_\mu\gamma_\nu-\gamma_\nu\gamma_\mu)$, and $\gamma_\mu$
and $\gamma_5$ are Dirac matrices. $q\equiv p_1-p_2$ is the four-momentum transfer, and $f_i(q^2)$ and $g_i(q^2)$ are the
vector and axial-vector form factors, respectively. Each form factor is assumed to be real in this work. The
four-momenta and masses of the particles involved in (\ref{eq:e1}) are $p_1=(E_1,{\mathbf p}_1)$ and $M_1$,
$p_2=(E_2,{\mathbf p}_2)$ and $M_2$, $l=(E,{\mathbf l})$ and $m$, and $p_\nu=(E_\nu^0,{\mathbf p}_\nu)$ and $m_\nu$,
respectively. Our calculations will be specialized to
the center-of-mass frame of $A$. In this frame, $p_1$, $p_2$, $l$, and $p_\nu$ will also represent the magnitudes of the corresponding
three-momenta, no confusion will arise from this. The directions of these momenta will be indicated by a caret, e.g.,
$\hat {\mathbf p}_2$.

Our approach to virtual RC follows the procedure of
Ref.~\cite{sirlin}. It has been discussed extensively in our
previous works (see Ref. \cite{tun91}), so only a few salient
facts will be repeated here. The virtual RC can be separated into
a model-independent part ${\mathsf M}_v$ and into a
model-dependent one which amounts to six constants. These latter
can be absorbed into the corresponding form factors of
(\ref{eq:e3}), this is indicated by a prime on ${\mathsf M}_0$.
The RC in ${\mathsf M}_v$ are finite in the ultraviolet, contain
the infrared cutoff, and are gauge invariant. We shall limit
ourselves here to exhibit explicitly only the new contributions of
order $(\alpha/\pi)(q/M_1)$ to the Dalitz plot with the ${\hat
{\mathbf s}_1} \cdot {\hat {\mathbf p}}_2$ correlation. However,
previous results are needed in the complete result. We shall give
ample referencing to them.

The transition amplitude with virtual RC is
\begin{equation}
{\mathsf M}_V = {\mathsf M}_0^\prime + {\mathsf M}_v.
\label{eq:a1}
\end{equation}
The calculation of all the integrals over the virtual photon
four-momentum that appear  in ${\mathsf M}_V$ have been performed
already to order $(\alpha/\pi)(q/M_1)$ in Ref.~\cite{tun91} for the
CDB case and in Ref.~\cite{tun93} for the NDB case. The
corresponding results are compactly expressed as
\begin{equation}
{\mathsf M}_{v\,i} = \frac{\alpha}{2\pi} \left[ {\mathsf M}_0
\Phi_i + {\mathsf M}_{{\not p}i} \Phi_i^\prime \right],
\label{eq:a2a}
\end{equation}
where $i=C,N$ separates the CDB and NDB cases, respectively. The
matrix element ${\mathsf M}_{{\not p}C}$ and the explicit forms of
$\Phi_C$ and $\Phi_C^\prime$ are found in Eqs.~(8), (6), and (7) of
Ref.~\cite{tun91}, respectively. The corresponding ones of
${\mathsf M}_{{\not p}N}$, $\Phi_N$, and $\Phi_N^\prime$ are found
in Eqs.~(6), (7), and (8) of Ref.~\cite{tun93}, once the
identifications $\Phi_N= 2{\mathrm Re\,} \phi$ and $\Phi_N^\prime
= 2{\mathrm Re\,}m\phi^\prime$ are made.

The Dalitz plot with virtual radiative corrections is now obtained
by leaving $E$ and $E_2$ as the relevant variables in the
differential decay rate for process (\ref{eq:e1}) and specializing
the result to exhibit explicitly the angular correlation ${\hat
{\mathbf s}_1} \cdot {\hat {\mathbf p}}_2$. After making the
replacement $u_A(p_1) \rightarrow \Sigma(s_1) u_A(p_1)$ in
${\mathsf M}_V$ [where the spin projector $\Sigma(s_1)$ is given
by $\Sigma(s_1) = (1 - \gamma_5 {\not \! s_1})/2$], squaring the
resulting amplitude, and rearranging terms we obtain for the
differential decay rate
\begin{equation}
d\Gamma_{iV} = d\Omega \left\{A_0^\prime + \frac{\alpha}{\pi}
(B_1^\prime \Phi_i + B_{i1}^{\prime \prime} \Phi_i^\prime) - {\hat
{\mathbf s}_1} \cdot {\hat {\mathbf p}_2} \left[ A_0^{\prime
\prime} + \frac{\alpha}{\pi} (B_2^\prime \Phi_i + B_{i2}^{\prime
\prime} \Phi_i^\prime) \right] \right\}, \label{eq:evc}
\end{equation}
In Eq.~(\ref{eq:evc}) the first two terms within curly brackets
correspond to the unpolarized Dalitz plot. For $i=C$ they can be
found in Ref.~\cite{tun91} where $A_0^\prime$, $B_1^\prime$, and
$B_{C1}^{\prime \prime}$, correspond to Eqs.~(10), (11), and (12),
respectively, of this reference. For $i=N$ the unpolarized Dalitz
plot can be found in Ref.~\cite{tun93}, where $B_{N1}^{\prime
\prime}$ corresponds to Eq.~(15). The spin-dependent part of
Eq.~(\ref{eq:evc}) was obtained to order $(\alpha/\pi)(q/M_1)^0$ in
Ref.~\cite{rfm97}. There in Eq.~(19) one can find the full
expression for $A_0^{\prime \prime}$. To the next order of
approximation, however, there appear the new contributions,
namely,
\begin{eqnarray}
B_2^\prime & = & Ep_2 \tilde Q_6 + Ely_0 \tilde Q_7, \label{eq:e5}
\end{eqnarray}
\begin{eqnarray}
B_{C2}^{\prime \prime} & = & Ep_2 Q_8 + Ely_0 Q_9, \label{eq:e6}
\end{eqnarray}
and
\begin{equation}
B_{N2}^{\prime \prime} = M_1p_2 Q_{N8} + M_1ly_0 Q_{N9}. \label{eq:e8}
\end{equation}
The phase space factor of Eq.~(\ref{eq:evc}) is $d\Omega =
({G_V^2}/{2})[{dE_2 \, dE\,d\Omega_2d\varphi_l}/{(2\pi)^5}]2M_1$,
the cosine $y_0$ of the angle between the directions of the
emitted baryon and the charged lepton is $ y_0 = [{(E_\nu^0)^2 -
l^2 - p_2^2}]/({2p_2l})$, and the neutrino energy is, by energy
conservation, $E_\nu^0 = M_1 - E_2 - E$.

The $Q_i$ in Eqs.~(\ref{eq:e5})-(\ref{eq:e8})are functions of
quadratic products of the form factors. They are new and are
listed in Appendix~A.

\section{Bremsstrahlung Radiative Corrections and final results}\label{sec:bremss}

The radiative process that accompanies (\ref{eq:e1}) is
\begin{equation}
A \to B \; \ell \; \overline{\nu}_l \; \gamma, \label{eq:e9}
\end{equation}
where the real photon $\gamma$ carries four-momentum $k =
(\omega,{\mathbf k})$ and the neutrino energy is now $E_\nu =
E_\nu^0-\omega$.

The Dalitz plot for this four-body decay covers the three-body
region of (\ref{eq:e1}) and extends over it by a region where both
$E_\nu$ and $\omega$ are always non-zero simultaneously. We shall
refer to this extension as the four-body region. A detailed
discussion of these two regions as well as explicit expressions of
their boundaries in the $(E,E_2)$ plane are given in
Ref.~\cite{rfm97}. Even if experiments have no provision to detect
the real photons in (\ref{eq:e9}), a precise measurement of $E$
and $E_2$ still allows to discriminate against photons belonging
to the four-body region. We shall assume in this paper that this
is the case and shall restrict our calculations to the three-body
region of (\ref{eq:e9}).

In order to establish our notation and conventions and to make the
necessary connections with our previous work, we must briefly
review the derivation of the bremsstrahlung differential decay rate.
According to the Low theorem \cite{low}, the amplitude for process
(\ref{eq:e9}) with contributions of orders $1/k$ and $k^0$ depends
only on the form factors of the non-radiative amplitude
(\ref{eq:e2}) and on the static electromagnetic multipoles of the
particles involved. The model dependence included by the real
photon appears in new form factors which vanish at least linearly
with $k$. These latter contribute to orders
$(\alpha/\pi)(q/M_1)^2$ and higher to the differential decay rate.
Thus, within the approximations of this paper  this part of the RC
is model independent. The transition amplitude consists of the sum
of three terms, namely,
\begin{equation}
{\mathsf M}_{iB} =  {\mathsf M}_{iB_1} + {\mathsf M}_{iB_2} +
{\mathsf M}_{iB_3}. \label{eq:b6}
\end{equation}
As in Sec.~\ref{sec:virtual}, the subindex $i=C,N$ is used to
distinguish CDB and NDB cases, respectively. The detailed
expressions of ${\mathsf M}_{CB_1}$, ${\mathsf M}_{CB_2}$, and
${\mathsf M}_{CB_3}$ are found in Eqs.~(18), (19), and (20),
respectively, of Ref.~\cite{tun91} and of ${\mathsf M}_{NB_1}$,
${\mathsf M}_{NB_2}$, and ${\mathsf M}_{NB_3}$ are found in
Eqs.~(21), (22), and (23) of Ref.~\cite{tun93}, respectively. Using
the $\Sigma(s_1)$ projector in Eq.~(\ref{eq:b6}), squaring the
matrix element, performing the trace calculations, inserting the
appropriate phase space factor, and indicating the integrations
over the photon variables, the differential decay rate can be
compactly given as
\begin{equation}
d\Gamma_{iB} = d\Gamma_{iB}^\prime - d\Gamma_{iB}^{(s)}.
\label{eq:e18}
\end{equation}
The analytical result to order $(\alpha/\pi)(q/M_1)$ of the
unpolarized decay rate $d\Gamma_{iB}^\prime$ was calculated in
Ref.~\cite{tun91} for the CDB case and in  Ref.~\cite{tun93} for the NDB
case. They can be found in Eqs.~(48) and (54) of such references,
respectively. The polarized decay rate $d\Gamma_{iB}^{(s)}$ was
calculated analytically to order $(\alpha/\pi)(q/M_1)^0$ in
Ref.~\cite{rfm97}. The final results are given in Eq.~(101) of this
reference, for both CDB and NDB cases.

The calculation to order $(\alpha/\pi)(q/M_1)$ of
$d\Gamma_{iB}^{(s)}$ is new. Let us now proceed with it. This
decay rate consists of the sum of three terms
\begin{equation}
d\Gamma_{iB}^{(s)} =
d\Gamma_{iBI}^{(s)}+d\Gamma_{iBII}^{(s)}+d\Gamma_{iBIII}^{(s)}.
\label{eq:b30}
\end{equation}
$d\Gamma_{iBI}^{(s)}$ comes from the product ${\mathsf
M}_{iB1}^{(s)}\overline{{\mathsf M}}_{iB1}$ and it contains the
infrared divergence and the finite terms that accompany it. To
extract them we follow the procedure used in Ref.~\cite{rfm97},
which extended the formalism introduced in Ref.~\cite{gins} for
$K_{l3}$ decays. The second and third summands in
Eq.~(\ref{eq:b30}) come from of the product $[{\mathsf
M}_{iB1}^{(s)} + {\mathsf M}_{iB2}^{(s)}] \overline{{\mathsf
M}}_{iBj}$ for $j=2,3$ respectively. They are infrared convergent
and are computed with standard techniques. The product ${\mathsf
M}_{iB3}^{(s)}\overline{{\mathsf M}}_{iB3}$ is left out because it
contributes to orders $(\alpha/\pi)(q/M_1)^2$ and higher. The
upper index $s$ indicates where the $\gamma_5 {\not \! s_1}$ of
$\Sigma(s_1)$ is contained in these amplitudes.

An important remark is in order here. It turns out that trying to
compute the terms of order $(\alpha/\pi)(q/M_1)$ only and then
adding them to the results of Ref.~\cite{rfm97} is long and more
cumbersome than doing from the start the full calculation
containing both $(\alpha/\pi)(q/M_1)^0$ and $(\alpha/\pi)(q/M_1)$
contributions. Accordingly, our new expressions will contain the
previous and the new contributions. It is then easy to verify that
by eliminating the $(\alpha/\pi)(q/M_1)$ terms in the new
expressions one obtains the ones of Ref.~\cite{rfm97}.

The procedure to calculate the CDB and NDB cases differ
substantially. We shall deal with them successively in the next
two subsections

\subsection{Charged decaying baryon case}

The polarized radiative differential decay rate can be cast into
the form
\begin{eqnarray}
d\Gamma_{CB}^{(s)} = \frac{\alpha}{\pi} \, d\Omega \, {\hat
{\mathbf s}_1} \cdot {\hat {\mathbf p}}_2 (B_2^\prime I_{C0} +
C_A^{(s)}). \label{eq:e11}
\end{eqnarray}
$I_{C0}$ contains the infrared divergence and the finite terms
that accompany it. It was calculated already, its explicit form is
found in Eq.~(52) of Ref.~\cite{rfm97}. $B_2^{\prime}$ contains new
$(q/M_1)$ contributions. It coincides with Eq.~(7) of the virtual
RC.  $C_A^{(s)}$ consists of the sum of three terms, namely,
\begin{equation}
C_A^{(s)} = \sum_{R=I}^{III} C_R, \label{eq:e12}
\end{equation}
where
\begin{equation}
C_R = \frac{p_2l}{2\pi}\int_{-1}^{y_0} dy \int_{-1}^1 dx \int_0^{2\pi} d\varphi_k \frac{|{\mathsf M_R}|^2}{D}.
\label{eq:e13}
\end{equation}
The integrations over the photon three-momentum are to be
performed through the variables $y={\hat {\mathbf p}_2} \cdot
{\hat {\mathbf l}}$, $x={\hat {\mathbf l}} \cdot \hat {\mathbf
k}$, and the azimuthal angle $\varphi_k$ of ${\mathbf k}$. The
traces of the square of the matrix elements give
\begin{eqnarray}
|{\mathsf M_I}|^2 & = & \frac{\beta^2(1-x^2)}{(1-\beta x)^2} \frac{E}{2} \left[-\frac{D}{p_2} {\tilde Q}_7
+ {\hat {\mathbf k}} \cdot {\hat {\mathbf p}_2}{\tilde Q}_9 + \frac{p_2(E+lx-D)}{M_1E} Q_{10}
+ \frac{(1-\beta x)(p_2+2 ly)}{M_1}Q_{11} \right. \nonumber \\
&  & \mbox{} +\left. \frac{2ly(E_\nu^0 + lx)+Dp_2}{M_1E}Q_{12} + \frac{ly}{M_1} Q_{13} - \frac{Dp_2}{M_1E} Q_{14} \right],
\label{eq:e14}
\end{eqnarray}
\begin{eqnarray}
|{\mathsf M_{II}}|^2 & = & \frac{1}{1-\beta x} \left[ \frac{p_2}{2} {\tilde Q}_6 + \frac{ly}{2} {\tilde Q}_7
+ \frac{p_2}{2}R_1 {Q}_8 + \left[ \frac{{\hat {\mathbf k}} \cdot {\hat {\mathbf p}_2}}{2}[ (E_\nu-\omega)R_2+\beta \omega x]
+ \frac{ly}{2}R_1 \right]{Q}_9 \right. \nonumber\\
&  & \mbox{} + \frac{p_2}{2M_1}\left[-({\hat {\mathbf k}} \cdot {\mathbf p}_2+lx+2\omega)R_2+\frac{\omega}{E} (2lx-D)\right]
Q_{10} \nonumber \\
&  & \mbox{} + \frac{p_2}{2M_1}\left[[-{\hat {\mathbf k}} \cdot {\hat {\mathbf p}_2}(p_2+2ly+2\omega{\hat {\mathbf k}} \cdot
{\hat {\mathbf p}_2})+lx]R_2+\frac{2l\omega y}{p_2}(1-\beta x) \right] Q_{11} \nonumber \\
&  & \mbox{} + \frac{p_2}{M_1} \left[ {\hat {\mathbf k}} \cdot {\hat {\mathbf p}_2}(p_2+ly+\omega {\hat {\mathbf k}} \cdot
{\hat {\mathbf p}_2})R_2+\frac{\omega}{2Ep_2} \left[Dp_2+2ly(D-{\hat {\mathbf k}} \cdot {\mathbf p}_2)\right] \right]Q_{12}
\nonumber \\
&  & \mbox{} + \left. \frac{l\omega y}{2M_1}Q_{13} - \frac{Dp_2\omega}{2 M_1E}Q_{14}-\frac{E_\nu}{2}{\hat {\mathbf k}} \cdot
{\hat {\mathbf p}_2} R_2 Q_{15} \right], \label{eq:e15}
\end{eqnarray}
and
\begin{eqnarray}
|{\mathsf M_{III}}|^2 & = & \frac{2E_\nu l}{M_1}\frac{(x {\hat {\mathbf k}} \cdot {\hat {\mathbf p}_2}-y)}{1-\beta x} Q_{16}
- \frac{l}{M_1}\frac{(x {\hat {\mathbf k}} \cdot {\hat {\mathbf p}_2}-y)}{1-\beta x} (E_\nu+\beta l+\beta p_2y+\beta \omega
x) Q_{17} \nonumber \\
&   & \mbox{} + \frac{E}{M_1} \frac{\beta^2(1-x^2)}{1-\beta x}(p_2+ly+\omega {\hat {\mathbf k}} \cdot {\hat {\mathbf p}_2})
Q_{18} \nonumber \\
&  & \mbox{} + \frac{l}{M_1} \left[\frac{{\hat {\mathbf k}} \cdot {\hat {\mathbf p}_2}}{1-\beta x}(\beta E_\nu-p_2y-l
-\omega x) + y \left[E_\nu+\frac{D-2E_\nu}{1-\beta x}\right] \right] Q_{19} \nonumber \\
&  & \mbox{} + \frac{l}{M_1} \left[\frac{{\hat {\mathbf k}} \cdot {\hat {\mathbf p}_2}}{1-\beta x}(\beta E_\nu+p_2y+l
+\omega x) + y \left[E_\nu-\frac{D}{1-\beta x}\right] \right] Q_{20} \nonumber \\
&  & \mbox{} - \frac{l}{M_1} \frac{\beta y[x(E_\nu^0-D)+p_2y+l]}{1-\beta x} Q_{21} \nonumber \\
&  & \mbox{} - \frac{\omega}{2M_1} \frac{{\hat {\mathbf k}} \cdot {\hat {\mathbf p}_2}}{1-\beta x}(E_\nu-D+\beta l+\beta p_2y
+ \beta \omega x) Q_{22} \nonumber \\
&  & \mbox{} + \frac{\omega}{2M_1}\frac{1}{1-\beta x} \left[{\hat {\mathbf k}} \cdot {\hat {\mathbf p}_2}(E_\nu-\beta l
- \beta p_2y-\beta \omega x)+\beta y(D-2E_\nu) \right]Q_{23} \nonumber \\
&  & \mbox{} + \frac{E_\nu \omega}{2M_1} {\hat {\mathbf k}} \cdot {\hat {\mathbf p}_2} Q_{24}-\frac{\omega}{2M_1}
(p_2+ly+\omega {\hat {\mathbf k}} \cdot {\hat {\mathbf p}_2})Q_{25}. \label{eq:e16}
\end{eqnarray}
Here $\beta=l/E$, $R_1 = -1+{\beta^2(1-x^2)}/({1-\beta
x})+{\omega}/{E}$, $R_2 = -1+({1-\beta^2})/({1-\beta
x})-{\omega}/{E}$, $D = E_\nu^0 + ({\mathbf l} + {\mathbf p}_2)
\cdot {\hat {\mathbf k}}$, and $\omega = {F}/({2D})$, with $F =
2p_2l(y_0-y)$.

The form factors of the vertex (\ref{eq:e3}) are contained in the
$Q_i$ coefficients. These are collected in Appendix \ref{appa}.

The complete differential decay rate, containing the Dalitz plot
with virtual and bremsstrahlung RC to order $(\alpha/\pi)(q/M_1)$,
is compactly expressed as
\begin{equation}
d\Gamma_{C} = d\Gamma_{CV} + d\Gamma_{CB}, \label{eq:e17}
\end{equation}
where the detailed expressions of $ d\Gamma_{CV}$ and
$d\Gamma_{CB}$, containing ${\hat s}_1$, can be traced starting
at Eqs.~(\ref{eq:evc}) and (\ref{eq:e11}). One can check that the
infrared cutoff $\lambda$ contained in the virtual RC is canceled
by its counterpart in the bremsstrahlung RC. Eq.~(\ref{eq:e17}) is
model-independent to order $(\alpha/\pi)(q/M_1)$. The photon
triple integrals of Eq.~({\ref{eq:e13}}) remain to be performed
numerically. This is our first main result, in the sense that it
can already be used in a Monte Carlo simulation. It complies with
all the requirements discussed in the Introduction to solve the
difficulties of the first two levels. However, it still presents
problems of the third level. The triple numerical integration form
is still unpractical. This difficulty can be substantially solved
because such triple integrations can be calculated analytically.

We shall now proceed to obtain the analytical counterpart of the
${\hat {\mathbf s}_1} \cdot {\hat {\mathbf p}}_2$ correlation
contained in Eq.~(\ref{eq:e17}). Within our approximations all the
form factors are constant and can be factored out of the very many
triple integrals. A convenient rearrangement of the $C_R$ of
Eq.~(\ref{eq:e12}) is
\begin{eqnarray}
C_I & = & \sum_{i=1}^8 Q_{i+6} \Lambda_i, \label{eq:e34} \\
C_{II} & = & \sum_{i=6}^{15} Q_i \Lambda_{i+3}, \label{eq:e35}
\end{eqnarray}
and
\begin{equation}
 C_{III}  =  \sum_{i=16}^{25} Q_i \Lambda_{i+3}.
\label{eq:e36}
\end{equation}
The $Q_i$ are the quadratic functions of  the form factors listed
in Appendix \ref{appa}.  The triple integrals are contained in the
$\Lambda_i$. We shall not detail here their explicit form in terms
of such integrals. We only give their final analytical expressions
and collect them in Appendix \ref{appb}. Many of these integrals
have been performed already in our previous work, although some
are new. To help the reader interested in following our
calculations in more detail, we have given in Appendix \ref{appd}
the general form of the triple integrals and a guide to identify
their analytical counterparts in our previous work. Only the
results for the new integrals are explicitly given in this
appendix. In organizing Eq.~(\ref{eq:e34}) with one running index
$i$ it was necessary to introduce $\Lambda_2=0$, because $Q_8$
does not appear in this equation.

The completely analytical result for the Dalitz plot in the
differential decay rate of CDB, Eq.~(\ref{eq:e17}), can be
compactly written as
\begin{equation}
d\Gamma_C = d\Omega \left[A_0^\prime-A_0^{\prime \prime} {\hat
{\mathbf s}_1} \cdot {\hat {\mathbf p}}_2 + \frac{\alpha}{\pi}
\left({\Theta_{CI}-\Theta_{CII}}{\hat {\mathbf s}_1} \cdot {\hat
{\mathbf p}}_2 \right) \right], \label{eq:e79}
\end{equation}
where
\begin{equation}
\Theta_{CI} = B_1^\prime (\Phi_C+I_{C0})+B_{C1}^{\prime
\prime}\Phi_C^\prime+C_A^\prime, \label{eq:e80}
\end{equation}
and
\begin{equation}
\Theta_{CII} = B_2^\prime (\Phi_C+I_{C0})+B_{C2}^{\prime\prime}
\Phi_C^\prime+C_A^{(s)}. \label{eq:e81}
\end{equation}

In this last equation $\Phi_C$, $\Phi_C^\prime$, $B_2^\prime$ and
$B_{C2}^{\prime\prime}$ are the same of Eq.~(\ref{eq:evc}),
$I_{C0}$ is the one of Eq.~(\ref{eq:e11}), and $C_A^{(s)}$ is given
by the sum of Eqs.~(\ref{eq:e34})-(\ref{eq:e36}). In
Eq.~(\ref{eq:e79}) $A_0^\prime$ and $A_0^{\prime \prime}$ are the
ones of Eq.~(\ref{eq:evc}) and the analytic form of $\Theta_{CI}$
is found in Eq.~(48) of Ref.~\cite{tun91}. One can check that when
$(q/M_1)$ contributions in $\Theta_{CII}$ and $\Theta_{CI}$ are
neglected, one obtains the order $(q/M_1)^0$ result of
Ref.~\cite{rfm97}. In particular $B_2^\prime$,
$B_{C2}^{\prime\prime}$, and $C_A^{(s)}$ become $A_2^\prime$,
$A_2^{\prime \prime}$, and $D_3(\rho_1+\rho_3) + D_4 (\rho_2 +
\rho_4)$ of Eqs.~(20), (21), and (96), respectively, of this
reference. Let us now proceed with the second case.

\subsection{Neutral decaying baryon case}

The calculation of $d\Gamma_{NB}^{(s)}$ proceeds in two ways. One
possibility is to perform a straight-forward calculation using the
tools described in the previous section. Another possibility is to
use the approach introduced in Ref.~\cite{tun93} to deal with the
convergent pieces of $d\Gamma_{NB}$. All the $(\alpha/\pi)(q/M_1)$
terms can be obtained using the approximation
\begin{equation}
\frac{1}{p_2 \cdot k} \simeq \frac{1}{p_1\cdot k} + \frac{q\cdot
k}{(p_1\cdot k)^2}, \label{eq:b57}
\end{equation}
and this will allow us to incorporate all the terms of order
$(\alpha/\pi)(q/M_1)$ that arise from this ratio. The advantage
 of this second possibility is that all the convergent terms of the NDB case are then obtained from their
counterparts for the CDB case up to a few additional terms. This
approximation, however, cannot be used in the divergent terms and
hence we need standard techniques to calculate them. In the first
term of Eq.~(\ref{eq:b30}) the infrared divergence is handled as
in Ref.~\cite{tun93} and afterwards the approximation
(\ref{eq:b57}) is used. One gets
\begin{equation}
d\Gamma_{NBI}^{(s)} = \frac{\alpha}{\pi} \, d\Omega \, {\hat
{\mathbf s}_1} \cdot {\hat {\mathbf p}_2} (B_2^\prime I_{N0}
+ C_I + {\tilde C}_{I}^{(s)}). \label{eq:b58}
\end{equation}
The next term in Eq.~(\ref{eq:b30}) can be arranged using
(\ref{eq:b57}) into
\begin{equation}
d\Gamma_{NBII}^{(s)} = \frac{\alpha}{\pi} \, d\Omega \, {\hat
{\mathbf s}_1} \cdot {\hat {\mathbf p}_2} (C_{II} + {\tilde
C}_{II}^{(s)}). \label{eq:b59}
\end{equation}
To calculate the third term in (\ref{eq:b30}) we can use the
approximations $W_\lambda \simeq \gamma_\lambda(f_1 +g_2\gamma_5)$
and $p_2 \simeq p_1$. The traces that arise are practically the
same as in $d\Gamma_{CBIII}^{(s)}$. However, there is a difference
to order $(\alpha/\pi)(q/M_1)$ which gives rise to a ${\tilde
C}_{III}^{(s)}$ summand. Thus, one gets
\begin{equation}
d\Gamma_{NBIII}^{(s)} = \frac{\alpha}{\pi} \, d\Omega \, {\hat
{\mathbf s}_1} \cdot {\hat {\mathbf p}_2} (C_{III} + {\tilde
C}_{III}^{(s)}). \label{eq:b60}
\end{equation}

The polarized decay rate becomes
\begin{equation}
d\Gamma_{NB}^{(s)} = \frac{\alpha}{\pi} \, d\Omega \, {\hat
{\mathbf s}_1} \cdot {\hat {\mathbf p}_2} (B_2^\prime I_{N0} +
C_A^{(s)} + C_{NA}^{(s)}). \label{eq:b61}
\end{equation}
$I_{N0}$ contains the infrared divergence and the finite terms
that accompany it. It was calculated already, its explicit form is
found in Eq.~(40) of Ref.~\cite{tun93}. As before in
Eq.~(\ref{eq:e5}), $B_2^\prime$ contains the $(\alpha/\pi)(q/M_1)$
contributions. $C_A^{(s)}$ is the same of Eq.~(\ref{eq:e11}) of
the CDB case and $C_{NA}^{(s)}$ is defined as
\begin{equation}
C_{NA}^{(s)} = {\tilde C}_I^{(s)} + {\tilde C}_{II}^{(s)} + {\tilde C}_{III}^{(s)}. \label{eq:e19}
\end{equation}

After some tedious but straight-forward trace calculations their
explicit forms are obtained. ${\tilde C}_i^{(s)}$ becomes
\begin{equation}
{\tilde C}_i^{(s)} = D_3\rho_i + D_4\rho_i^\prime. \label{eq:e20}
\end{equation}
with $i=I,II,III$. Here $D_3=2(-g_1^2+f_1 g_1)$ and
$D_4=2(g_1^2+f_1 g_1)$. $\rho_i$ and $\rho_i^\prime$ are
\begin{equation}
\rho_I = \frac{p_2l}{2\pi M_1} \int_{-1}^{y_0} dy \int_{-1}^1 dx \int_0^{2\pi} d\varphi_k\frac{\beta [-y+x
{\hat {\mathbf p}}_2 \cdot {\hat {\mathbf k}}]}{D (1-\beta x)}(DE_\nu^0 + p_2ly), \label{eq:e21}
\end{equation}
\begin{equation}
\rho_I^\prime = \frac{p_2l}{2\pi M_1} \int_{-1}^{y_0} dy
\int_{-1}^1 dx \int_0^{2\pi} d\varphi_k \frac{l[-y+x {\hat
{\mathbf p}}_2 \cdot {\hat {\mathbf k}}]}{D(1-\beta x)}[-D + {\hat
{\mathbf k}} \cdot {\mathbf p}_2], \label{eq:e22}
\end{equation}
\begin{equation}
\rho_{II} = \frac{lp_2^2}{8\pi M_1} \int_{-1}^{y_0} dy \int_{-1}^1 dx \int_0^{2\pi} d\varphi_k \frac{E_\nu}{D} \left[
1 + \frac{\beta y-{\hat {\mathbf p}}_2 \cdot {\hat {\mathbf k}}}{1-\beta x} \, {\hat {\mathbf k}} \cdot {\hat {\mathbf p}_2}
\right], \label{eq:e24}
\end{equation}
\begin{equation}
\rho_{II}^\prime = \frac{lp_2^2}{8\pi M_1}\int_{-1}^{y_0} dy
\int_{-1}^1 dx \int_0^{2\pi} d\varphi_k  \frac{E_\nu}{D}\left[
{\hat {\mathbf p}}_2 \cdot {\hat {\mathbf k}}+\frac{\beta y-{\hat
{\mathbf p}}_2 \cdot {\hat {\mathbf k}}} {1-\beta x}\right] {\hat
{\mathbf p}_\nu} \cdot {\hat {\mathbf p}_2}, \label{eq:e25}
\end{equation}
\begin{eqnarray}
\rho_{III} & = & \frac{p_2l}{4\pi M_1}\int_{-1}^{y_0}dy\int_{-1}^1 dx \int_0^{2\pi} d\varphi_k \frac{E_\nu E}{D}\left\{\left[
\frac{1-\beta^2}{1-\beta x}-\frac{2\omega}{E}-1\right] {\hat {\mathbf k}} \cdot {\hat {\mathbf p}_2} \right. \nonumber \\
&  & \mbox{} - \left. \beta y\left[ 1 - \beta{\hat {\mathbf p}}_\nu \cdot \frac{x{\hat {\mathbf k}}-{\hat {\mathbf l}}}
{1-\beta x}\right] \right\}, \label{eq:e27}
\end{eqnarray}
and
\begin{eqnarray}
\rho_{III}^\prime & = & \frac{p_2l}{4\pi M_1}\int_{-1}^{y_0}dy\int_{-1}^1 dx \int_0^{2\pi} d\varphi_k \frac{EE_\nu}{D}\left\{
\left[ \frac{1-\beta^2}{1-\beta x}-(1+\beta x)-\frac{2\omega}{E}\right] \right. {\hat {\mathbf p}_\nu} \cdot
{\hat {\mathbf p}_2} \nonumber \\
&  & \mbox{} - \left. \beta y\left[ \frac{1-{\hat {\mathbf p}}_\nu \cdot {\hat {\mathbf k}}}{1-\beta x} \right] + \left[
\frac{1-\beta {\hat {\mathbf l}} \cdot {\hat {\mathbf p}}_\nu}{1-\beta x}-1\right] {\hat {\mathbf k}} \cdot
{\hat {\mathbf p}_2} \right\}, \label{eq:e28}
\end{eqnarray}

The complete differential decay rate that contains the Dalitz plot
of the NDB case including the ${\hat {\mathbf s}_1} \cdot {\hat
{\mathbf p}_2}$ correlation can be expressed compactly as
\begin{equation}
d\Gamma_N = d\Gamma_{NV} + d\Gamma_{NB}, \label{eq:e31}
\end{equation}
where the detailed expressions of $d\Gamma_{NV}$ and
$d\Gamma_{NB}$ containing ${\hat s}_1$ are traced starting at
Eqs.~(\ref{eq:evc}) and (\ref{eq:b61}). This is our second main
result and the discussion of the previous subsection applies to
it: its triple numerical integration form is still unpractical.
This difficulty is solved by performing analytically the triple
integrals contained in Eq.~(\ref{eq:b61}). We have to concentrate
only on $C_{NA}^{(s)}$ of this equation, all other terms in it have
been given analytically already. Then, Eq.~(\ref{eq:e19}) can be
cast into the compact form
\begin{equation}
C_{NA}^{(s)} = D_3 \rho_{N3} + D_4 \rho_{N4}, \label{eq:e98}
\end{equation}
where
\begin{equation}
\rho_{N3} = \rho_I+\rho_{II}+\rho_{III}, \label{eq:e99}
\end{equation}
and
\begin{equation}
\rho_{N4} = \rho_I^\prime+\rho_{II}^\prime+\rho_{III}^\prime.
\label{eq:e100}
\end{equation}
The explicit analytical expressions of the $\rho_i$ and
$\rho_i^\prime$ are collected in Appendix \ref{appc}.

The completely analytical result for the Dalitz plot in
Eq.~(\ref{eq:e31}) can be put in parallel with Eq.~(\ref{eq:e79}),
namely,
\begin{equation}
d\Gamma_N = d\Omega \left[A_0^\prime-A_0^{\prime \prime}
{\hat{\mathbf s}_1} \cdot {\hat {\mathbf p}}_2 +
\frac{\alpha}{\pi} \left({\Theta_{NI}-\Theta_{NII}}{\hat {\mathbf
s}_1} \cdot {\hat{\mathbf p}}_2\right) \right], \label{eq:e95}
\end{equation}
where
\begin{equation}
\Theta_{NI} = B_1^\prime (\Phi_N+I_{N0}) + B_{N1}^{\prime \prime}
\Phi_N^\prime + C_A^\prime + C_{NA}^\prime, \label{eq:e96}
\end{equation}
and
\begin{equation}
\Theta_{NII} = B_2^\prime (\Phi_N+I_{N0}) + B_{N2}^{\prime \prime}
\Phi_N^\prime + C_A^{(s)} + C_{NA}^{(s)}. \label{eq:e97}
\end{equation}

In this last equation $\Phi_N$, $\Phi_N^\prime$, $B_{2}^\prime$
and $B_{N2}^{\prime\prime}$ are the same of Eq.~(\ref{eq:evc}),
$C_A^{(s)}$ and $I_{N0}$ are the ones of Eq.~(\ref{eq:b61}), and
$C_{NA}^{(s)}$ is given in Eq.~(\ref{eq:e98}). In Eq.~(\ref{eq:e95})
$A_0^\prime$ and $A_0^{\prime \prime}$ are the ones of
Eq.~(\ref{eq:evc}) and the analytical form of $\Theta_{NI}$ is
found in Eq.~(54) of Ref.~\cite{tun93}. One can check that when
$(\alpha/\pi)(q/M_1)$ contributions in Eq.~(\ref{eq:e95}) are
neglected one obtains the $(\alpha/\pi)(q/M_1)^0$ result of
Ref.~\cite{rfm97}. In particular $B_{N2}^{\prime\prime}$ becomes
$A_2^{\prime\prime}$ of this reference and $C_{NA}^{(s)}$ becomes
zero.

\section{Crosschecks}\label{sec:cks}

There are several points we want to make in this section. One is
that the analytical results are so long that it is important to
check them. Another one is that there are some results already
available in the literature \cite{toth} and we should compare with
them. An even more important point is to provide the reader
interested in using our result with numbers to be reproduced.

To crosscheck the analytical results we use the triple numerical
integration form of the RC. We make numerical comparisons of both
forms by fixing the values of several form factors and of the
Dalitz plot variables $E$ and $E_2$. A complete crosscheck
requires the use of several choices of non-zero values for all the
six form factors and a range of values of the pair $(E,E_2)$ over
the Dalitz plot. Also, the comparison with the numerical results
of Ref.~\cite{toth} should be made in the several cases covered
there. All these crosschecks and comparisons were satisfactory and
it is not necessary to display all the details here. Accordingly,
we shall present a minimum of numerical tables and limit our
discussion to them.

For definiteness, we shall work with the decays $\Sigma^- \to n e
{\overline \nu}$ and $\Lambda \to p e {\overline \nu}$ as examples
of CDB and NDB cases. The reason for this is that numerical RC for
these two decays were produced in Ref.~\cite{toth}. We shall
accordingly fix the form factors at the values used in this
reference, namely, $g_1/f_1 =-0.34$, $f_2/f_1=-0.97$ for $\Sigma^-
\to n e {\overline \nu}$ decay and $g_1/f_1 =0.72$, $f_2/f_1=0.97$
for $\Lambda \to p e {\overline \nu}$. In addition, we use $f_1=1$
in $\Sigma^- \to n e {\overline \nu}$ and $f_1=1.2366$ in $\Lambda
\to p e {\overline \nu}$ and to compare with Ref.~\cite{toth} in
both these decays we put $g_2=0$ and neglect $f_3$ and $g_3$
contributions. The values of the masses come from
Ref.~\cite{part}. The anomalous magnetic moments of the baryons
appear in our expressions of the RC. We use $\kappa(\Sigma^-)=0.3764M_N$, $\kappa(\Lambda)=0.6130M_N$,
$\kappa(n)=1.9130M_N$, and $\kappa(p)=-1.7928M_N$,
where $M_N$ is the nuclear magneton. These values are extracted from
the corresponding total magnetic moments reported in \cite{part}
using Eq.~(22) of Ref.~\cite{tun91}. We neglected the anomalous
magnetic moment of the electron, due to its smallness.

As an example of the numerical crosscheck we display Table I for
$\Sigma^- \to n e {\overline \nu}$, where for generality we
allowed $g_2,g_3,f_3\neq 0$. In the upper entries (a) we use the
triple numerical integration form to obtain the RC for $C_A^{(s)}$
of the ${\hat {\mathbf s}_1} \cdot {\hat {\mathbf p}}_2$
correlation covering a lattice of points over the Dalitz plot. The
energies $E$ and $E_2$ enter through $\delta=E/E_m$ and
$\sigma=E_2/M_1$. $E_m$, $\sigma^{max}$, and $\sigma^{min}$ are
determined using the boundaries of the three body region given in
Ref.~\cite{rfm97}. The lower entries (b) contain the RC for the
same $C_A^{(s)}$ calculated with the analytical form. An
inspection of this table shows an agreement to two decimal places
and the third one being close.

\begingroup
\squeezetable
\begin{table}
\begin{tabular}{lrrrrrrrrrr}
\hline\hline $\sigma$ &
\multicolumn{10}{c}{(a)} \\
\hline
 0.8077 &$-$0.0744 &$-$0.0811 &$-$0.0587 &$-$0.0256 &   0.0101 &   0.0420 &   0.0650 &   0.0741 &   0.0648 &   0.0309 \\
 0.8056 &$-$0.1350 &$-$0.1435 &$-$0.1048 &$-$0.0507 &   0.0049 &   0.0528 &   0.0852 &   0.0955 &   0.0778 &   0.0253 \\
 0.8035 &          &$-$0.1673 &$-$0.1294 &$-$0.0708 &$-$0.0090 &   0.0443 &   0.0802 &   0.0910 &   0.0700 &   0.0102 \\
 0.8014 &          &$-$0.1875 &$-$0.1534 &$-$0.0921 &$-$0.0254 &   0.0327 &   0.0718 &   0.0831 &   0.0590 &          \\
 0.7993 &          &$-$0.2065 &$-$0.1784 &$-$0.1152 &$-$0.0438 &   0.0193 &   0.0617 &   0.0734 &   0.0459 &          \\
 0.7972 &          &          &$-$0.2054 &$-$0.1407 &$-$0.0643 &   0.0042 &   0.0503 &   0.0623 &   0.0306 &          \\
 0.7951 &          &          &$-$0.2358 &$-$0.1697 &$-$0.0873 &$-$0.0124 &   0.0379 &   0.0499 &   0.0128 &          \\
 0.7930 &          &          &$-$0.2720 &$-$0.2042 &$-$0.1141 &$-$0.0309 &   0.0246 &   0.0363 &          &          \\
 0.7909 &          &          &          &$-$0.2480 &$-$0.1469 &$-$0.0520 &   0.0106 &   0.0211 &          &          \\
 0.7888 &          &          &          &$-$0.3115 &$-$0.1913 &$-$0.0772 &$-$0.0035 &   0.0041 &          &          \\
 0.7867 &          &          &          &          &$-$0.2684 &$-$0.1110 &$-$0.0135 &          &          &          \\ \\
        & \multicolumn{10}{c}{(b)} \\ \hline
 0.8077 &$-$0.0744 &$-$0.0810 &$-$0.0587 &$-$0.0256 &   0.0098 &   0.0412 &   0.0639 &   0.0732 &   0.0643 &   0.0309 \\
 0.8056 &$-$0.1350 &$-$0.1435 &$-$0.1048 &$-$0.0507 &   0.0047 &   0.0521 &   0.0842 &   0.0947 &   0.0774 &   0.0253 \\
 0.8035 &          &$-$0.1673 &$-$0.1293 &$-$0.0706 &$-$0.0092 &   0.0437 &   0.0793 &   0.0902 &   0.0697 &   0.0102 \\
 0.8014 &          &$-$0.1875 &$-$0.1533 &$-$0.0919 &$-$0.0255 &   0.0322 &   0.0710 &   0.0823 &   0.0587 &          \\
 0.7993 &          &$-$0.2065 &$-$0.1783 &$-$0.1149 &$-$0.0437 &   0.0189 &   0.0609 &   0.0727 &   0.0457 &          \\
 0.7972 &          &          &$-$0.2053 &$-$0.1404 &$-$0.0640 &   0.0040 &   0.0497 &   0.0617 &   0.0305 &          \\
 0.7951 &          &          &$-$0.2357 &$-$0.1693 &$-$0.0869 &$-$0.0124 &   0.0374 &   0.0495 &   0.0127 &          \\
 0.7930 &          &          &$-$0.2719 &$-$0.2036 &$-$0.1135 &$-$0.0308 &   0.0243 &   0.0360 &          &          \\
 0.7909 &          &          &          &$-$0.2473 &$-$0.1460 &$-$0.0516 &   0.0104 &   0.0210 &          &          \\
 0.7888 &          &          &          &$-$0.3106 &$-$0.1900 &$-$0.0765 &$-$0.0034 &   0.0041 &          &          \\
 0.7867 &          &          &          &          &$-$0.2666 &$-$0.1100 &$-$0.0134 &          &          &          \\
 \hline
$\delta$& 0.0500 & 0.1500 & 0.2500 & 0.3500 & 0.4500 & 0.5500 & 0.6500 & 0.7500 & 0.8500 & 0.9500 \\ \\
$\sigma^{max}$& 0.8078 & 0.8078 & 0.8078 & 0.8078 & 0.8078 & 0.8078 & 0.8078 & 0.8078 & 0.8078 & 0.8078 \\
$\sigma^{min}$& 0.8043 & 0.7978 & 0.7925 & 0.7884 & 0.7857 & 0.7847 & 0.7854 & 0.7884 & 0.7939 & 0.8023 \\
\hline\hline
\end{tabular}
\caption{Values of $C_A^{(s)}$ in $\Sigma^- \to n e {\overline
\nu}$ decay by (a) integrating it numerically and (b) evaluating
it analytically. $C_A^{(s)}$ is given in units of
$\textrm{GeV}^2$. The form factors have been given the arbitrary
values $f_1=1.0$, $f_2 = -0.97$, $f_3 = -0.778$, $g_1 = -0.34$,
$g_2 = 0.987$, and $g_3 = -1.563$. }
\end{table}
\endgroup

To proceed with the comparison with Ref.~\cite{toth} we must use
the difference defined there, namely,
\begin{equation}
\delta \alpha_B(E,E_2) = \alpha_B(E,E_2) - \alpha_0(E,E_2), \label{eq:e101}
\end{equation}
where
\begin{equation}
\alpha_{Bi}(E,E_2) = -\frac{A_0^{\prime\prime}(E,E_2) +
(\alpha/\pi) \Theta_{iII}(E,E_2)}{A_0^\prime(E,E_2) + (\alpha/\pi)
\Theta_{iI}(E,E_2)}, \label{eq:e102}
\end{equation}
and as before $i=C,N$ and $\alpha_0(E,E_2) =
-{A_0^{\prime\prime}(E,E_2)}/{A_0^\prime(E,E_2)}$ for both values
of $i$.

One may interpret $\alpha_{Bi}(E,E_2)$ as the asymmetry parameter
of the emitted baryon at $(E,E_2)$ points of the Dalitz plot. Here
we must choose the same $(E,E_2)$ points as in Ref.~\cite{toth}
and, as already mentioned, use the same values of the form
factors. However, it should be stressed that our final results are
not compromised to fixing the values of the form factors.

Before proceeding with a detailed comparison with the numbers of
this reference, there is a point that must be kept in mind. The
approximations used in our work and in Ref.~\cite{toth} are not
quite the same. We used the Low theorem to calculate the
bremsstrahlung RC and in this reference it was assumed that both
baryons involved were point-like and higher
$(\alpha/\pi)(q/M_1)^n$ contributions ($n\geq 2$) were included in
this part. Another interesting thing is to compare our order
$(\alpha/\pi)(q/M_1)$ results with our previous order
$(\alpha/\pi)(q/M_1)^0$ results. As explained earlier these latter
are reproduced here when the $(\alpha/\pi)(q/M_1)$ contributions
are neglected.

We performed many comparisons and, as before, there is no need to
present all the details. One example, the $\Lambda \to p e
{\overline \nu}$ case is enough for this discussion. The results
are displayed in Table II. In the upper (a) part only the order
$(\alpha/\pi)(q/M_1)^0$ is given. Both this order and the order
$(\alpha/\pi)(q/M_1)$ contributions are added in the middle part
(b). The numerical results of Ref.~\cite{toth} are reproduced in
the lower part (c). A numerical crosscheck was also performed in
producing parts (a) and (b). We do not reproduce it here, the
agreement was as good as in Table I.

An inspection of Table II shows that the order
$(\alpha/\pi)(q/M_1)$ is systematically perceptible at the second
significant digit and even at the first one. In comparing with
Ref.~\cite{toth}, one can see a better agreement with the middle
table (b). The agreement at the first significant digit improves
as the RC grow in size and also the variations in the second digit
become smaller. There are differences, however. They may be
explained as due to the different approximations used. Also,
comparing entries (a) and (b) one may conclude that for light
quark hyperon semileptonic decays the order $(\alpha/\pi)(q/M_1)$
is perceptible enough and that when heavy quarks are involved
contributions of this order become relevant in precision
experiments.

\begingroup
\squeezetable
\begin{table}
\begin{tabular}{lrrrrrrrrrr}
\hline\hline
$\sigma$ &
\multicolumn{10}{c}{(a)} \\
\hline
 0.8530 & 0.1 & 0.0 & 0.0 & 0.0 & 0.0 & 0.1 & 0.1 & 0.2 & 0.4 & 1.4 \\
 0.8518 & 1.2 & 0.2 & 0.1 & 0.1 & 0.2 & 0.2 & 0.3 & 0.5 & 1.0 & 1.9 \\
 0.8505 &     & 0.5 & 0.3 & 0.3 & 0.3 & 0.4 & 0.6 & 0.9 & 1.5 &     \\
 0.8492 &     & 1.0 & 0.5 & 0.5 & 0.5 & 0.6 & 0.8 & 1.2 & 1.6 &     \\
 0.8480 &     & 1.7 & 0.9 & 0.7 & 0.7 & 0.8 & 1.1 & 1.4 & 1.4 &     \\
 0.8467 &     &     & 1.3 & 1.0 & 1.0 & 1.1 & 1.3 & 1.5 & 0.5 &     \\
 0.8454 &     &     & 1.9 & 1.4 & 1.3 & 1.4 & 1.5 & 1.5 &     &     \\
 0.8442 &     &     &     & 1.9 & 1.7 & 1.7 & 1.7 & 1.0 &     &     \\
 0.8429 &     &     &     & 2.7 & 2.3 & 2.1 & 1.6 &     &     &     \\
 0.8416 &     &     &     &     & 3.4 & 2.4 & 0.1 &     &     &     \\ \\
        & \multicolumn{10}{c}{(b)} \\ \hline
 0.8530 & 0.5 & 0.2 & 0.2 & 0.2 & 0.1 & 0.1 & 0.1 & 0.2 & 0.3 & 0.9 \\
 0.8518 & 2.0 & 0.5 & 0.4 & 0.4 & 0.4 & 0.4 & 0.5 & 0.6 & 0.9 & 1.1 \\
 0.8505 &     & 0.9 & 0.7 & 0.6 & 0.6 & 0.7 & 0.8 & 1.0 & 1.3 &     \\
 0.8492 &     & 1.5 & 1.0 & 0.9 & 0.9 & 0.9 & 1.1 & 1.3 & 1.3 &     \\
 0.8480 &     & 2.3 & 1.3 & 1.1 & 1.1 & 1.2 & 1.3 & 1.4 & 1.0 &     \\
 0.8467 &     &     & 1.8 & 1.5 & 1.4 & 1.5 & 1.5 & 1.5 & 0.2 &     \\
 0.8454 &     &     & 2.3 & 1.8 & 1.7 & 1.7 & 1.7 & 1.3 &     &     \\
 0.8442 &     &     &     & 2.2 & 2.0 & 1.9 & 1.7 & 0.7 &     &     \\
 0.8429 &     &     &     & 2.8 & 2.4 & 2.1 & 1.5 &     &     &     \\
 0.8416 &     &     &     &     & 3.2 & 2.2 & 0.1 &     &     &     \\ \\
        & \multicolumn{10}{c}{(c)} \\ \hline
 0.8530 & 0.2 & 0.1 & 0.1 & 0.1 & 0.1 & 0.1 & 0.1 & 0.2 & 0.3 & 0.9 \\
 0.8518 & 1.9 & 0.4 & 0.3 & 0.3 & 0.3 & 0.4 & 0.5 & 0.6 & 1.0 & 1.1 \\
 0.8505 &     & 0.8 & 0.6 & 0.5 & 0.6 & 0.6 & 0.8 & 1.0 & 1.3 &     \\
 0.8493 &     & 1.4 & 0.9 & 0.8 & 0.8 & 0.9 & 1.1 & 1.3 & 1.4 &     \\
 0.8480 &     & 2.4 & 1.3 & 1.1 & 1.1 & 1.2 & 1.3 & 1.5 & 1.1 &     \\
 0.8467 &     &     & 1.7 & 1.4 & 1.4 & 1.5 & 1.6 & 1.5 & 0.3 &     \\
 0.8455 &     &     & 2.3 & 1.8 & 1.7 & 1.7 & 1.7 & 1.3 &     &     \\
 0.8442 &     &     &     & 2.2 & 2.0 & 2.0 & 1.7 & 0.8 &     &     \\
 0.8429 &     &     &     & 2.8 & 2.5 & 2.2 & 1.5 &     &     &     \\
 0.8417 &     &     &     &     & 3.2 & 2.3 & 0.1 &     &     &     \\ \\
\hline
$\delta$& 0.0500 & 0.1500 & 0.2500 & 0.3500 & 0.4500 & 0.5500 & 0.6500 & 0.7500 & 0.8500 & 0.9500 \\ \\
$\sigma^{max}$& 0.8536 & 0.8536 & 0.8536 & 0.8536 & 0.8536 & 0.8536 & 0.8536 & 0.8536 & 0.8536 & 0.8536 \\
$\sigma^{min}$& 0.8516 & 0.8479 & 0.8450 & 0.8428 & 0.8414 & 0.8410 & 0.8416 & 0.8433 & 0.8464 & 0.8508 \\
\hline\hline
\end{tabular}
\caption{\label{table:comthl} $100\delta \alpha_B(E,E_2)$ with RC
over the three-body region in $\Lambda \to p e {\overline \nu}$
decay. (a) gives the RC to order $(\alpha/\pi)(q/M_1)^0$, (b)
gives the RC to order $(\alpha/\pi)(q/M_1)$, and (c) corresponds
to the RC computed in Ref.~\cite{toth}.}
\end{table}
\endgroup

Let us now turn to a different form to use our results. A form
which may provide a more efficient use of them in a Monte Carlo
simulation and which still is not compromised to fixing values of
the form factors, as was the case in Table II.

\section{Numerical form of the Radiative Corrections}\label{sec:numerical}

We now come to our second goal in this paper. In the previous
sections we have obtained the RC to CDB and NDB in two forms. The
first one has triple integrals over the real photon variables
ready to be performed numerically. The second one is fully
analytical. Although this latter one is already practical it is
still long and tedious. It still requires that the RC be
calculated within the Monte Carlo simulation every time $E$ and
$E_2$ are varied. This is much faster than performing the triple
integrals, but, it still represents a non-negligible computer
effort. We shall now discuss a third form of the RC that may be
more practical to use.

For fixed values of $E$ and $E_2$, Eqs.~(\ref{eq:e80}) and
(\ref{eq:e81}) for the CDB case and Eqs.~(\ref{eq:e96}) and
(\ref{eq:e97}) for the NDB case take the form
\begin{equation}
\Theta_m = \sum_{i\leq j=1}^6 a_{ij}^m f_i f_j, \label{eq:e105}
\end{equation}
because they are quadratic in the form factors. For the sake of
simplify, in Eq.~(\ref{eq:e105}) we have momentarily redefined
$g_1=f_4$, $g_2 =f_5$, and $g_3=f_6$. Notice that the restriction
$i\leq j$ reduces the sum in Eq.~(\ref{eq:e105}) to 21 terms. The
index $m$ takes the values $m=CI$, $CII$, $NI$, and $NII$. The
third form of RC we propose consists of calculating arrays of the
$a_{ij}^m$ coefficients determined at fixed values of $(E,E_2)$
and that these pairs of $(E,E_2)$ cover a lattice of points on the
Dalitz plot.

To calculate the coefficients $a_{ij}^m$ it is not necessary to
rearrange our final results, either analytical or to be
integrated, so that they take the form (\ref{eq:e105}). One can
calculate them following a systematic procedure. One chooses fixed
$(E,E_2)$ points. Then one fixes $f_1=1$ and $f_i=0$, $i\neq 1$
and obtains $a_{11}^m$, one repeats this calculation for $f_2=1$,
$f_i=0$, $i\neq 2$ to obtain $a_{22}^m$, and again until $f_6=1$,
$f_i=0$, $i\neq 6$ and $a_{66}^m$ is obtained. Next, one repeats
the calculation with $f_1=1$, $f_2=1$, $f_i=0$, $i\neq 1,2$ and
from this results one subtracts $a_{11}^m$ and $a_{22}^m$, this
way one obtains the coefficient $a_{12}^m$. One repeats this last
step changing $i$ and $j$ until all the interference coefficients
$a_{ij}^m$, $i\neq j$, have been calculated.

To illustrate all this and to further discuss it we have produced
arrays presented in two tables, selecting in each one ten points
$(E,E_2)$ over the Dalitz plot. We have chosen two examples,
$\Lambda \to p e {\overline \nu}$ of a NDB case which is displayed
in Table III and $\Lambda_c^+ \to \Lambda e^+\nu$ of a CDB case
which is displayed in Table IV. This latter also serves as an
example of a heavy quark decay. As in the previous section, the
more important purpose is to provide the user with numbers to
compare with. The arrays of these two tables were obtained using
the RC in the analytical form. In the $\Lambda_c^+$ case we used
the formulas for the charge assignments $A^-$, $B^0$, $l^-$ of the
CDB case of the previous sections and then applied the rules of
Ref.~\cite{rfm02} to obtain the results for the charge assignment
$A^+$, $B^0$, $l^+$ of this particular case. In these tables we
have restored our standard notation for the axial-vector form
factors $g_1$, $g_2$, and $g_3$. The masses used are those of
Sec.~\ref{sec:cks}, $M_1(\Lambda_c^+)$ comes from Ref.~\cite{part},
and we assume an estimate for $\kappa(\Lambda_c^+)=0.1106M_N$.

\begin{turnpage}

\begingroup
\squeezetable
\begin{table}
\begin{tabular}{lrrrrrrrrrr}
\hline
    & (0.05,0.8518) & (0.35,0.8518) & (0.65,0.8518) & (0.95,0.8518) & (0.25,0.8480) & (0.55,0.8480) & (0.75,0.8480) & (0.45,0.8442) & (0.65,0.8442) & (0.55,0.8416) \\
\hline
$f_1^2 $ & $ 6.812\times 10^{-4}$ & $ 1.550\times 10^{-4}$ & $-1.024\times 10^{-4}$ & $ 6.505\times 10^{-4}$ & $ 7.218\times 10^{-4}$ & $-1.622\times 10^{-4}$ & $-6.987\times 10^{-5}$ & $ 7.198\times 10^{-5}$ & $-1.227\times 10^{-4}$ & $-4.887\times 10^{-5}$ \\
$f_2^2 $ & $ 1.604\times 10^{-3}$ & $-1.875\times 10^{-4}$ & $-5.732\times 10^{-4}$ & $ 2.204\times 10^{-3}$ & $ 6.787\times 10^{-4}$ & $-1.268\times 10^{-3}$ & $-3.588\times 10^{-4}$ & $-7.888\times 10^{-4}$ & $-5.807\times 10^{-4}$ & $-2.633\times 10^{-4}$ \\
$f_3^2 $ & $ 0.000              $ & $ 0.000$               & $ 0.000$               & $ 0.000$               & $ 0.000$               & $ 0.000$               & $ 0.000$               & $ 0.000$               & $ 0.000$               & $ 0.000$ \\
$g_1^2 $ & $ 7.470\times 10^{-2}$ & $-8.900\times 10^{-3}$ & $-2.635\times 10^{-2}$ & $ 1.026\times 10^{-1}$ & $ 4.812\times 10^{-2}$ & $-9.116\times 10^{-2}$ & $-2.537\times 10^{-2}$ & $-1.250\times 10^{-1}$ & $-9.181\times 10^{-2}$ & $-2.073\times 10^{-1}$ \\
$g_2^2 $ & $ 1.882\times 10^{-3}$ & $-2.199\times 10^{-4}$ & $-6.722\times 10^{-4}$ & $ 2.585\times 10^{-3}$ & $ 1.227\times 10^{-3}$ & $-2.293\times 10^{-3}$ & $-6.488\times 10^{-4}$ & $-3.131\times 10^{-3}$ & $-2.305\times 10^{-3}$ & $-5.219\times 10^{-3}$ \\
$g_3^2 $ & $ 0.000$               & $ 0.000$               & $ 0.000$               & $ 0.000$               & $ 0.000$               & $ 0.000$               & $ 0.000$               & $ 0.000$               & $ 0.000$               & $ 0.000$ \\
$f_1f_2$ & $ 2.010\times 10^{-3}$ & $-4.892\times 10^{-5}$ & $-5.756\times 10^{-4}$ & $ 2.395\times 10^{-3}$ & $ 1.258\times 10^{-3}$ & $-1.173\times 10^{-3}$ & $-3.495\times 10^{-4}$ & $-4.985\times 10^{-4}$ & $-5.694\times 10^{-4}$ & $-2.477\times 10^{-4}$ \\
$f_1f_3$ & $ 0.000$               & $ 0.000$               & $ 0.000$               & $ 0.000$               & $ 0.000$               & $ 0.000$               & $ 0.000$               & $ 0.000$               & $ 0.000$               & $ 0.000$ \\
$f_2f_3$ & $ 0.000$               & $ 0.000$               & $ 0.000$               & $ 0.000$               & $ 0.000$               & $ 0.000$               & $ 0.000$               & $ 0.000$               & $ 0.000$               & $ 0.000$ \\
$g_1g_2$ & $-2.362\times 10^{-2}$ & $ 2.889\times 10^{-3}$ & $ 8.504\times 10^{-3}$ & $-3.252\times 10^{-2}$ & $-1.509\times 10^{-2}$ & $ 2.906\times 10^{-2}$ & $ 8.219\times 10^{-3}$ & $ 3.975\times 10^{-2}$ & $ 2.911\times 10^{-2}$ & $ 6.574\times 10^{-2}$ \\
$g_1g_3$ & $ 0.000$               & $ 0.000$               & $ 0.000$               & $ 0.000$               & $ 0.000$               & $ 0.000$               & $ 0.000$               & $ 0.000$               & $ 0.000$               & $ 0.000$ \\
$g_2g_3$ & $ 0.000$               & $ 0.000$               & $ 0.000$               & $ 0.000$               & $ 0.000$               & $ 0.000$               & $ 0.000$               & $ 0.000$               & $ 0.000$               & $ 0.000$ \\
$f_1g_1$ & $ 1.683\times 10^{-1}$ & $ 3.604\times 10^{-1}$ & $ 1.875\times 10^{-2}$ & $-5.329\times 10^{-2}$ & $ 2.878\times 10^{-1}$ & $ 8.435\times 10^{-2}$ & $-1.250\times 10^{-1}$ & $ 1.258\times 10^{-1}$ & $-8.629\times 10^{-2}$ & $-3.368\times 10^{-2}$ \\
$f_1g_2$ & $-1.763\times 10^{-3}$ & $-6.857\times 10^{-3}$ & $ 6.870\times 10^{-4}$ & $-9.573\times 10^{-4}$ & $-1.808\times 10^{-2}$ & $-4.425\times 10^{-3}$ & $ 9.176\times 10^{-3}$ & $-1.410\times 10^{-2}$ & $ 1.046\times 10^{-2}$ & $ 5.085\times 10^{-3}$ \\
$f_1g_3$ & $ 0.000$               & $ 0.000$               & $ 0.000$               & $ 0.000$               & $ 0.000$               & $ 0.000$               & $ 0.000$               & $ 0.000$               & $ 0.000$               & $ 0.000$ \\
$f_2g_1$ & $-2.482\times 10^{-2}$ & $-8.236\times 10^{-3}$ & $-6.172\times 10^{-3}$ & $ 2.989\times 10^{-2}$ & $-2.846\times 10^{-2}$ & $-1.758\times 10^{-2}$ & $-4.052\times 10^{-3}$ & $-1.703\times 10^{-2}$ & $-9.910\times 10^{-3}$ & $-5.365\times 10^{-3}$ \\
$f_2g_2$ & $ 3.970\times 10^{-3}$ & $ 1.329\times 10^{-3}$ & $ 9.192\times 10^{-4}$ & $-4.778\times 10^{-3}$ & $ 4.634\times 10^{-3}$ & $ 2.719\times 10^{-3}$ & $ 4.868\times 10^{-4}$ & $ 2.659\times 10^{-3}$ & $ 1.401\times 10^{-3}$ & $ 7.390\times 10^{-4}$ \\
$f_2g_3$ & $ 0.000$               & $ 0.000$               & $ 0.000$               & $ 0.000$               & $ 0.000$               & $ 0.000$               & $ 0.000$               & $ 0.000$               & $ 0.000$               & $ 0.000$ \\
$f_3g_1$ & $-5.014\times 10^{-5}$ & $-9.348\times 10^{-5}$ & $-6.324\times 10^{-5}$ & $-5.962\times 10^{-7}$ & $-1.323\times 10^{-4}$ & $-1.003\times 10^{-4}$ & $-2.914\times 10^{-5}$ & $-9.289\times 10^{-5}$ & $-2.412\times 10^{-5}$ & $-9.643\times 10^{-6}$ \\
$f_3g_2$ & $-4.271\times 10^{-5}$ & $-8.147\times 10^{-5}$ & $-5.522\times 10^{-5}$ & $-5.513\times 10^{-7}$ & $-1.136\times 10^{-4}$ & $-8.644\times 10^{-5}$ & $-2.517\times 10^{-5}$ & $-7.876\times 10^{-5}$ & $-2.061\times 10^{-5}$ & $-8.140\times 10^{-6}$ \\
$f_3g_3$ & $ 0.000$               & $ 0.000$               & $ 0.000$               & $ 0.000$               & $ 0.000$               & $ 0.000$               & $ 0.000$               & $ 0.000$               & $ 0.000$               & $ 0.000$ \\
\hline
\end{tabular}
\caption{Numerical arrays of the coefficients $a_{ij}^{NII}$ in
$\textrm{GeV}^2$ of Eq.~(\ref{eq:e105}) evaluated at ten points
$(E,E_2)$ (headings of columns) over the polarized Dalitz plot of
$\Lambda \to p e^- \bar{\nu}$ decay.}
\end{table}
\endgroup

\end{turnpage}

\begin{turnpage}

\begingroup
\squeezetable
\begin{table}
\begin{center}
\begin{tabular}{
l r@{.}l@{}r r@{.}l@{}r r@{.}l@{}r r@{.}l@{}r r@{.}l@{}r
r@{.}l@{}r r@{.}l@{}r r@{.}l@{}r r@{.}l@{}r r@{.}l@{}r }
\hline\hline & \multicolumn{3}{c}{(0.15,0.5995)} &
\multicolumn{3}{c}{(0.45,0.5995)} &
\multicolumn{3}{c}{(0.75,0.5995)} &
\multicolumn{3}{c}{(0.95,0.5995)} &
\multicolumn{3}{c}{(0.25,0.5602)} &
\multicolumn{3}{c}{(0.55,0.5602)} &
\multicolumn{3}{c}{(0.85,0.5602)} &
\multicolumn{3}{c}{(0.45,0.5210)} &
\multicolumn{3}{c}{(0.75,0.5210)} &
\multicolumn{3}{c}{(0.65,0.4948)}
\\ \hline
$f_1^2 $ & $-1$ & 532 & $\times 10^{-2}$ & $ 1$ & 036 & $\times 10^{-1}$ & $ 6$ & 859 & $\times 10^{-3}$ & $-3$ & 894 & $\times 10^{-1}$ & $ 7$ & 582 & $\times 10^{-2}$ & $ 2$ & 298 & $\times 10^{-1}$ & $-3$ & 036 & $\times 10^{-1}$ & $ 2$ & 295 & $\times 10^{-1}$ & $-3$ & 851 & $\times 10^{-2}$ & $ 6$ & 446 & $\times 10^{-2}$ \\
$f_2^2 $ & $-4$ & 818 & $\times 10^{-3}$ & $ 2$ & 436 & $\times 10^{-1}$ & $ 1$ & 849 & $\times 10^{-2}$ & $-8$ & 624 & $\times 10^{-1}$ & $ 2$ & 289 & $\times 10^{-1}$ & $ 5$ & 389 & $\times 10^{-1}$ & $-6$ & 710 & $\times 10^{-1}$ & $ 5$ & 611 & $\times 10^{-1}$ & $-8$ & 135 & $\times 10^{-2}$ & $ 1$ & 481 & $\times 10^{-1}$ \\
$f_3^2 $ & $-0$ & 000 & \multicolumn{1}{c}{} & $-0$ & 000 & \multicolumn{1}{c}{} & $-0$ & 000 & \multicolumn{1}{c}{} & $-0$ & 000 & \multicolumn{1}{c}{} & $-0$ & 000 & \multicolumn{1}{c}{} & $-0$ & 000 & \multicolumn{1}{c}{} & $-0$ & 000 & \multicolumn{1}{c}{} & $-0$ & 000 & \multicolumn{1}{c}{} & $-0$ & 000 & \multicolumn{1}{c}{} & $-0$ & 000 & \multicolumn{1}{c}{} \\
$g_1^2 $ & $-1$ & 466 & $\times 10^{-2}$ & $ 1$ & 076 & \multicolumn{1}{c}{} & $ 8$ & 152 & $\times 10^{-2}$ & $-3$ & 808 & \multicolumn{1}{c}{} & $ 1$ & 534 & \multicolumn{1}{c}{} & $ 3$ & 546 & \multicolumn{1}{c}{} & $-4$ & 416 & \multicolumn{1}{c}{} & $ 7$ & 830 & \multicolumn{1}{c}{} & $-1$ & 139 & \multicolumn{1}{c}{} & $ 1$ & 004 & $\times 10^{+1}$ \\
$g_2^2 $ & $-5$ & 571 & $\times 10^{-3}$ & $ 2$ & 817 & $\times 10^{-1}$ & $ 2$ & 138 & $\times 10^{-2}$ & $-9$ & 972 & $\times 10^{-1}$ & $ 3$ & 943 & $\times 10^{-1}$ & $ 9$ & 282 & $\times 10^{-1}$ & $-1$ & 156 & \multicolumn{1}{c}{} & $ 2$ & 047 & \multicolumn{1}{c}{} & $-2$ & 967 & $\times 10^{-1}$ & $ 2$ & 632 & \multicolumn{1}{c}{} \\
$g_3^2 $ & $-0$ & 000 & \multicolumn{1}{c}{} & $-0$ & 000 & \multicolumn{1}{c}{} & $-0$ & 000 & \multicolumn{1}{c}{} & $-0$ & 000 & \multicolumn{1}{c}{} & $-0$ & 000 & \multicolumn{1}{c}{} & $-0$ & 000 & \multicolumn{1}{c}{} & $-0$ & 000 & \multicolumn{1}{c}{} & $-0$ & 000 & \multicolumn{1}{c}{} & $-0$ & 000 & \multicolumn{1}{c}{} & $-0$ & 000 & \multicolumn{1}{c}{} \\
$f_1f_2$ & $-3$ & 445 & $\times 10^{-2}$ & $ 3$ & 133 & $\times 10^{-1}$ & $ 2$ & 160 & $\times 10^{-2}$ & $-1$ & 159 & \multicolumn{1}{c}{} & $ 2$ & 497 & $\times 10^{-1}$ & $ 6$ & 956 & $\times 10^{-1}$ & $-9$ & 031 & $\times 10^{-1}$ & $ 7$ & 047 & $\times 10^{-1}$ & $-1$ & 130 & $\times 10^{-1}$ & $ 1$ & 941 & $\times 10^{-1}$ \\
$f_1f_3$ & $ 0$ & 000 & \multicolumn{1}{c}{} & $ 0$ & 000 & \multicolumn{1}{c}{} & $ 0$ & 000 & \multicolumn{1}{c}{} & $ 0$ & 000 & \multicolumn{1}{c}{} & $ 0$ & 000 & \multicolumn{1}{c}{} & $ 0$ & 000 & \multicolumn{1}{c}{} & $ 0$ & 000 & \multicolumn{1}{c}{} & $ 0$ & 000 & \multicolumn{1}{c}{} & $ 0$ & 000 & \multicolumn{1}{c}{} & $ 0$ & 000 & \multicolumn{1}{c}{} \\
$f_2f_3$ & $ 0$ & 000 & \multicolumn{1}{c}{} & $ 0$ & 000 & \multicolumn{1}{c}{} & $ 0$ & 000 & \multicolumn{1}{c}{} & $ 0$ & 000 & \multicolumn{1}{c}{} & $ 0$ & 000 & \multicolumn{1}{c}{} & $ 0$ & 000 & \multicolumn{1}{c}{} & $ 0$ & 000 & \multicolumn{1}{c}{} & $ 0$ & 000 & \multicolumn{1}{c}{} & $ 0$ & 000 & \multicolumn{1}{c}{} & $ 0$ & 000 & \multicolumn{1}{c}{} \\
$g_1g_2$ & $ 8$ & 296 & $\times 10^{-3}$ & $-1$ & 114 & \multicolumn{1}{c}{} & $-8$ & 738 & $\times 10^{-2}$ & $ 3$ & 897 & \multicolumn{1}{c}{} & $-1$ & 571 & \multicolumn{1}{c}{} & $-3$ & 659 & \multicolumn{1}{c}{} & $ 4$ & 514 & \multicolumn{1}{c}{} & $-8$ & 037 & \multicolumn{1}{c}{} & $ 1$ & 150 & \multicolumn{1}{c}{} & $-1$ & 030 & $\times 10^{+1}$ \\
$g_1g_3$ & $ 0$ & 000 & \multicolumn{1}{c}{} & $ 0$ & 000 & \multicolumn{1}{c}{} & $ 0$ & 000 & \multicolumn{1}{c}{} & $ 0$ & 000 & \multicolumn{1}{c}{} & $ 0$ & 000 & \multicolumn{1}{c}{} & $ 0$ & 000 & \multicolumn{1}{c}{} & $ 0$ & 000 & \multicolumn{1}{c}{} & $ 0$ & 000 & \multicolumn{1}{c}{} & $ 0$ & 000 & \multicolumn{1}{c}{} & $ 0$ & 000 & \multicolumn{1}{c}{} \\
$g_2g_3$ & $ 0$ & 000 & \multicolumn{1}{c}{} & $ 0$ & 000 & \multicolumn{1}{c}{} & $ 0$ & 000 & \multicolumn{1}{c}{} & $ 0$ & 000 & \multicolumn{1}{c}{} & $ 0$ & 000 & \multicolumn{1}{c}{} & $ 0$ & 000 & \multicolumn{1}{c}{} & $ 0$ & 000 & \multicolumn{1}{c}{} & $ 0$ & 000 & \multicolumn{1}{c}{} & $ 0$ & 000 & \multicolumn{1}{c}{} & $ 0$ & 000 & \multicolumn{1}{c}{} \\
$f_1g_1$ & $ 9$ & 235 & \multicolumn{1}{c}{} & $ 9$ & 436 & $\times 10^{-1}$ & $-9$ & 728 & \multicolumn{1}{c}{} & $-2$ & 410 & \multicolumn{1}{c}{} & $ 8$ & 070 & \multicolumn{1}{c}{} & $-2$ & 512 & \multicolumn{1}{c}{} & $-5$ & 571 & \multicolumn{1}{c}{} & $ 5$ & 378 & \multicolumn{1}{c}{} & $-6$ & 121 & \multicolumn{1}{c}{} & $-2$ & 391 & \multicolumn{1}{c}{} \\
$f_1g_2$ & $-4$ & 429 & $\times 10^{-1}$ & $ 8$ & 400 & $\times 10^{-2}$ & $ 7$ & 031 & $\times 10^{-1}$ & $-8$ & 759 & $\times 10^{-1}$ & $-1$ & 671 & \multicolumn{1}{c}{} & $ 7$ & 110 & $\times 10^{-1}$ & $ 5$ & 766 & $\times 10^{-1}$ & $-2$ & 146 & \multicolumn{1}{c}{} & $ 2$ & 184 & \multicolumn{1}{c}{} & $ 1$ & 134 & \multicolumn{1}{c}{} \\
$f_1g_3$ & $ 0$ & 000 & \multicolumn{1}{c}{} & $ 0$ & 000 & \multicolumn{1}{c}{} & $ 0$ & 000 & \multicolumn{1}{c}{} & $ 0$ & 000 & \multicolumn{1}{c}{} & $ 0$ & 000 & \multicolumn{1}{c}{} & $ 0$ & 000 & \multicolumn{1}{c}{} & $ 0$ & 000 & \multicolumn{1}{c}{} & $ 0$ & 000 & \multicolumn{1}{c}{} & $ 0$ & 000 & \multicolumn{1}{c}{} & $ 0$ & 000 & \multicolumn{1}{c}{} \\
$f_2g_1$ & $-4$ & 903 & $\times 10^{-1}$ & $-2$ & 339 & $\times 10^{-1}$ & $ 1$ & 770 & $\times 10^{-2}$ & $ 3$ & 505 & \multicolumn{1}{c}{} & $-2$ & 921 & $\times 10^{-2}$ & $-6$ & 449 & $\times 10^{-1}$ & $ 3$ & 069 & \multicolumn{1}{c}{} & $ 1$ & 706 & \multicolumn{1}{c}{} & $ 7$ & 521 & $\times 10^{-1}$ & $ 2$ & 504 & $\times 10^{-1}$ \\
$f_2g_2$ & $ 3$ & 501 & $\times 10^{-1}$ & $ 1$ & 071 & $\times 10^{-1}$ & $-1$ & 408 & $\times 10^{-1}$ & $-1$ & 853 & \multicolumn{1}{c}{} & $ 2$ & 102 & $\times 10^{-1}$ & $ 2$ & 404 & $\times 10^{-1}$ & $-1$ & 777 & \multicolumn{1}{c}{} & $-7$ & 944 & $\times 10^{-1}$ & $-5$ & 591 & $\times 10^{-1}$ & $-1$ & 778 & $\times 10^{-1}$ \\
$f_2g_3$ & $ 0$ & 000 & \multicolumn{1}{c}{} & $ 0$ & 000 & \multicolumn{1}{c}{} & $ 0$ & 000 & \multicolumn{1}{c}{} & $ 0$ & 000 & \multicolumn{1}{c}{} & $ 0$ & 000 & \multicolumn{1}{c}{} & $ 0$ & 000 & \multicolumn{1}{c}{} & $ 0$ & 000 & \multicolumn{1}{c}{} & $ 0$ & 000 & \multicolumn{1}{c}{} & $ 0$ & 000 & \multicolumn{1}{c}{} & $ 0$ & 000 & \multicolumn{1}{c}{} \\
$f_3g_1$ & $-2$ & 067 & $\times 10^{-2}$ & $-2$ & 617 & $\times 10^{-2}$ & $-1$ & 189 & $\times 10^{-2}$ & $-4$ & 642 & $\times 10^{-4}$ & $-3$ & 808 & $\times 10^{-2}$ & $-3$ & 174 & $\times 10^{-2}$ & $-3$ & 158 & $\times 10^{-3}$ & $-3$ & 269 & $\times 10^{-2}$ & $-4$ & 636 & $\times 10^{-3}$ & $-3$ & 500 & $\times 10^{-3}$ \\
$f_3g_2$ & $-1$ & 261 & $\times 10^{-2}$ & $-1$ & 597 & $\times 10^{-2}$ & $-7$ & 253 & $\times 10^{-3}$ & $-2$ & 833 & $\times 10^{-4}$ & $-2$ & 147 & $\times 10^{-2}$ & $-1$ & 790 & $\times 10^{-2}$ & $-1$ & 781 & $\times 10^{-3}$ & $-1$ & 704 & $\times 10^{-2}$ & $-2$ & 417 & $\times 10^{-3}$ & $-1$ & 731 & $\times 10^{-3}$ \\
$f_3g_3$ & $ 0$ & 000 & \multicolumn{1}{c}{} & $ 0$ & 000 & \multicolumn{1}{c}{} & $ 0$ & 000 & \multicolumn{1}{c}{} & $ 0$ & 000 & \multicolumn{1}{c}{} & $ 0$ & 000 & \multicolumn{1}{c}{} & $ 0$ & 000 & \multicolumn{1}{c}{} & $ 0$ & 000 & \multicolumn{1}{c}{} & $ 0$ & 000 & \multicolumn{1}{c}{} & $ 0$ & 000 & \multicolumn{1}{c}{} & $ 0$ & 000 & \multicolumn{1}{c}{} \\
\hline
\end{tabular}
\caption{Numerical arrays of the coefficients $a_{ij}^{CII}$ in
$\textrm{GeV}^2$ of Eq.~(\ref{eq:e105}) evaluated at ten points
$(E,E_2)$ (headings of columns) over the polarized Dalitz plot of
$\Lambda_c^+ \to \Lambda e^+ \nu$ decay.}
\end{center}
\end{table}
\endgroup

\end{turnpage}

The first fact that appears in these tables is that the RC do not
depend on the form factor products $f_3^2$, $g_3^2$, $f_1f_3$,
$f_2f_3$, $g_1g_3$, $g_2g_3$, $f_1g_3$, $f_2g_3$, and $f_3g_3$ in
Tables III and IV. The non-appearance of these products cannot be
seen easily in our final results of Sec.~\ref{sec:bremss}. The
other fact is that the non-zero RC to each form factor product
vary appreciably from one $(E,E_2)$ point to another. This means
that replacing the precision results of Sec.~\ref{sec:bremss} with
an array of only a few columns over the Dalitz plot is far from
satisfactory. Therefore the lattice of $(E,E_2)$ points must be
much finer than only a few points.

The use of this third presentation of RC is very practical in the
sense that such RC can be calculated separately and only the
arrays should be fed into the Monte Carlo simulation. However, in
a precision experiment possibly involving 150, 200, and even 300
bins over the Dalitz plot the number of columns in the RC arrays
should be at least just as many. It may be required that several
columns be produced in finer subdivisions within each bin,
possibly 4, 8, or even more. For example, one may require that the
numerical changes of the $a_{ij}^m$ coefficients between
neighboring $(E,E_2)$ points do not exceed two decimal places
within rounding of the third decimal place.

To close this section let us stress that none of the three forms
of our RC results is compromised to fixing from the outset values
for the form factors when RC are applied in a Monte Carlo
simulation. To fix them at prescribed values may be not too bad an
assumption for hyperon semileptonic decays, but it is not
acceptable at all for decays involving heavy quarks where the
Cabibbo theory \cite{part} is no longer reliable for fixing the
form factors.

\section{Summary and conclusions}\label{sec:diss}

We have obtained in Secs.~\ref{sec:virtual} and
\ref{sec:bremss} the RC to the angular correlation ${\hat {\mathbf
s}_1} \cdot {\hat {\mathbf p}_2}$ to order $(\alpha/\pi)(q/M_1)$.
Our final results are given in two forms. The first one is the
triple numerical integration form, in which the integrations over
the real photon variables are explicitly exhibited and remain to
be performed numerically. The second one is the analytical form
where those integrations have all been calculated analytically. We
covered two cases, the CDB and the NDB ones whose final results
are given  in Eqs.~(\ref{eq:e79}) and (\ref{eq:e95}), respectively.
The analytical results are very long and tedious. To make their
use more accessible we have collected the numerous $Q_i$ and
$\Lambda_i$ algebraic expressions which appear in the CDB case in
Appendices \ref{appa} and \ref{appb}, respectively. The NDB case
uses these expressions and also the $\rho_i$ ones. These latter
were collected in Appendix \ref{appc}.

Our analytical results were crosschecked and compared with other
results available in literature. This we have done in
Sec.~\ref{sec:cks}. We have limited ourselves to discuss the decay
$\Sigma^- \to n e {\overline \nu}$ as an example of the
crosschecks and $\Lambda \to p e {\overline \nu}$ as an example of
the comparisons with Ref.~\cite{toth}. In addition, in this latter
decay we also included a comparison between our RC to orders
$(\alpha/\pi)(q/M_1)^0$ and $(\alpha/\pi)(q/M_1)$.

We have discussed in Sec.~\ref{sec:numerical} another possibility
to use our results in an experimental analysis. One can calculate
the numerical factors of the quadratic products of the form
factors that appear in the RC at fixed values of $(E,E_2)$. These
factors can be organized in arrays to be multiplied upon such
products, covering a lattice of $(E,E_2)$ points over the Dalitz
plot. We discussed two examples of this possibility, a CDB one and
NDB one.

Apart from illustration purposes, the several tables in this paper
provide numbers to compare with. Also, our calculations rely
heavily on previous results. Apart from discussions in text, we
have given in Appendix \ref{appd} details to allow the
identification of our previous and new analytical results for the
many integrals.

To close, let us recall that our results are general within our
approximation. They can be applied in the other four charge
assignments of baryons involving heavy quarks and whether the
charged lepton is $e^\pm$, $\mu^\pm$, or $\tau^\pm$. They are
model independent and are not compromised to fixing the form
factors at prescribed values. The above calculations should be
extended to cover precision RC in  the ${\hat {\mathbf s}_1} \cdot
{\hat {\mathbf l}}$ correlation \cite{mtz01} and in the four body region \cite{mtz02}. We
hope to return to these cases in the near future.

\acknowledgments The authors are grateful to Consejo Nacional de
Ciencia y Tecnolog{\'\i}a (M\'exico) for partial support.
J.J.~Torres, A.~M., and M.~Neri were partially supported by
Comisi\'on de Operaci\'on y Fomento de Actividades Acad\'emicas
(Instituto Polit\'ecnico Nacional). They also wish to thank the
warm hospitality extended to them at IF-UASLP, where part of this
paper was performed. R.~F.-M.\ was also partially supported by
Fondo de Apoyo a la Investigaci\'on (Universidad Aut\'onoma de San
Luis Potos{\'\i}).

\appendix

\section{Collection of the $Q_i$ coefficients}\label{appa}

The coefficients $Q_i$ introduced in Secs.~II and III are long
quadratic functions of the form factors. The coefficients
$Q_1,\ldots, Q_7$ have been computed in previous works
\cite{tun91,rfm97} and can be found there. The new coefficients
are listed below. They read
\begin{eqnarray*}
{\tilde Q}_6 & = & F_1^2 \left[ \frac{E_2 - M_2 - \beta p_2 y_0}{M_1} \right] + G_1^2 \left[
\frac{E_2 + M_2 - \beta p_2 y_0}{M_1} \right] + 2F_1G_1 \left[ \frac{E_2-\beta p_2 y_0}{M_1} \right] \nonumber \\
&  & \mbox{} - (F_1F_2-G_1G_2) \left[ \frac{\beta p_2 y_0}{M_1} \right] - F_1G_2 \left[ \frac{M_1-M_2+E_\nu^0-E}{M_1}
- \frac{q^2}{2M_1 E} \right] \nonumber \\
&  & \mbox{} + F_2G_1 \left[\frac{M_1+M_2+E_\nu^0-E}{M_1}-\frac{q^2}{2M_1 E} \right] - F_2G_2 \left[ \frac{2E_\nu^0}{M_1}
- \frac{q^2}{2M_1 E} \right],
\end{eqnarray*}
\begin{eqnarray*}
{\tilde Q}_7 & = & F_1^2 \left[1+\frac{M_2}{M_1}\right]\left[\frac{E_2 - M_2}{E} \right] + G_1^2 \left[ 1-\frac{M_2}{M_1}
\right] \left[\frac{E_2 + M_2}{E} \right] - 2 F_1 G_1 \left[ \frac{E_\nu^0-E}{E} \right] \nonumber \\
&  & \mbox{} + F_1 G_2 \left[\frac{E_2 - M_2}{M_1} \right] \left[ \frac{E_\nu^0-E}{E} \right]- F_2 G_1 \left[
\frac{E_2 + M_2}{M_1} \right] \left[ \frac{E_\nu^0-E}{E} \right] \nonumber \\
&  & \mbox{} + (F_1 F_2 - G_1 G_2) \left[ \frac{p_2^2}{M_1 E} \right],
\end{eqnarray*}
\begin{eqnarray*}
Q_8 & = & F_1^2 \left[\frac{E_2-M_2}{M_1} \right] + G_1^2 \left[\frac{E_2+M_2}{M_1} \right] + 2 F_1 G_1 \left[
\frac{E_2}{M_1} \right] + F_1G_2\left[\frac{E-M_1+M_2}{M_1} \right] \nonumber \\
&  & \mbox{} - F_2 G_2 \left[\frac{E_\nu^0}{M_1} \right]- F_2 G_1 \left[ \frac{E-M_1-M_2}{M_1} \right] + F_3 G_1 \left[
\frac{E(E_2+M_2)}{M_1^2}\right],
\end{eqnarray*}
\begin{eqnarray*}
Q_9 & = & F_1^2 \left[\frac{E_2-M_2}{M_1} \right] + G_1^2 \left[\frac{E_2+M_2}{M_1} \right] + 2 F_1 G_1 \left[
\frac{E_2}{M_1} \right]- F_1 G_2 \left[ \frac{E_2-M_2}{M_1} \right] \nonumber \\
&   & \mbox{} + F_2 G_1 \left[\frac{E_2+M_2}{M_1}\right] - F_3 G_1 \left[ \frac{E_2}{M_1}-1\right]\left[\frac{E_2+M_2}{M_1}
\right],
\end{eqnarray*}
\begin{eqnarray*}
Q_{N8} & = & F_1^2 \left[\frac{(M_1-E)(E_2-M_2)}{M_1^2} \right] + G_1^2 \left[ \frac{(M_1-E)(E_2+M_2)}{M_1^2}\right]
+ 2 F_1G_1 \left[ \frac{M_2^2}{M_1^2}-\frac{E_\nu^0}{M_1} \right] \nonumber \\
&  & \mbox{} + F_1 G_2 \left[ \frac{M_2(E_2-M_2-E_\nu^0)}{M_1^2} \right] + F_2 G_1 \left[ \frac{M_2(E_2+M_2-E_\nu^0)}{M_1^2}
\right] \nonumber \\
&  & \mbox{} - F_2G_2 \left[ \frac{E_2 E_\nu^0}{M_1^2} \right] + F_3G_1 \left[ \frac{E(E_2+M_2)}{M_1^2}\right],
\end{eqnarray*}
\begin{eqnarray*}
Q_{N9} & = & -F_1^2 \left[\frac{M_2(E_2-M_2)}{M_1^2} \right] + G_1^2 \left[\frac{M_2(E_2+M_2)}{M_1^2} \right]
+ 2 F_1G_1 \left[ \frac{M_2^2}{M_1^2} \right] - F_1 G_2 \left[\frac{E_2(E_2-M_2)}{M_1^2} \right] \nonumber \\
&  & \mbox{} + F_2 G_1 \left[ \frac{E_2(E_2+M_2)}{M_1^2} \right] + F_3G_1 \left[\frac{M_2+E_2}{M_1}\right]\left[1
- \frac{E_2}{M_1} \right],
\end{eqnarray*}
\[
Q_{10}=-F_2G_1+F_1G_2+F_2G_2,
\]
\[
Q_{11} = \left[\frac{E_2+M_2}{M_1} \right] G_1 F_3,
\]
\[
Q_{12} = 2 F_1 G_1,
\]
\begin{eqnarray*}
Q_{13} & = & -F_1^2 \left[ \frac{E_2-M_2}{E}\right] - G_1^2 \left[ \frac{E_2+M_2}{E} \right] + 2 F_1 G_1 \left[\frac{E_2}{E}
\right] + F_2G_1 \left[\frac{E_2+M_2}{E} \right] \nonumber \\
&  & \mbox{} - F_1G_2 \left[\frac{E_2-M_2}{E} \right] - F_3G_1 \left[1-\frac{E_2}{M_1} \right]\left[\frac{M_2+E_2}{E}\right],
\end{eqnarray*}
\[
Q_{14} = -F_1^2 - G_1^2 - F_1F_2 + G_1G_2,
\]
\[
Q_{15} = 2 F_1^2 \left[\frac{E_2-M_2}{M_1} \right] + 2 G_1^2 \left[\frac{E_2+M_2}{M_1} \right],
\]
\[
Q_{16} = f_1(g_2-g_1) - f_2g_1,
\]
\[
Q_{17} = f_1g_2+ f_3 g_1,
\]
\[
Q_{18} = \frac{1}{2}(f_1^2-g_1^2) + f_2(f_1+g_1) - g_1(f_3-g_2),
\]
\[
Q_{19} = 2 f_1 g_1 \left[\frac{1}{2 M_1}+ \frac{\kappa_1}{e} \right] M_1,
\]
\[
Q_{20} = -2 g_1^2 \left[\frac{1}{2 M_1}+ \frac{\kappa_1}{e} \right] M_1,
\]
\[
Q_{21} = \frac{1}{2}(f_1^2-g_1^2) + f_2(f_1-g_1) + g_1(f_3+g_2),
\]
\[
Q_{22} = (f_1-g_1)(f_2-g_2)+M_1\frac{\kappa_2}{e}(f_1-g_1)^2-M_1\frac{\kappa_1}{e}(f_1^2-g_1^2),
\]
\[
Q_{23} = -(f_1+g_1)(f_2-g_2)+M_1\frac{\kappa_1}{e}(f_1+g_1)^2-M_1\frac{\kappa_2}{e}(f_1^2-g_1^2),
\]
\begin{eqnarray*}
Q_{24} & = & -(f_1-g_1)^2+g_1(2 f_1+3f_2+2f_3+g_2-2g_1)-f_1(f_2+g_2) \nonumber \\
&  & \mbox{} - M_1 \frac{\kappa_1}{e} (f_1-g_1)^2 + M_1 \frac{\kappa_2}{e} (f_1^2-g_1^2) - 4 M_1 \frac{\kappa_1}{e} g_1^2,
\end{eqnarray*}
and
\begin{eqnarray*}
Q_{25} & = &-(f_1^2-g_1^2)-(f_1+g_1)(f_2+g_2)+2g_1(f_3-f_2) \nonumber \\
&  & \mbox{} + M_1 \frac{\kappa_2}{e} (5g_1^2+f_1^2+2f_1g_1)-M_1 \frac{\kappa_1}{e}(f_1^2-g_1^2).
\end{eqnarray*}
Here, $\kappa_1$ and $\kappa_2$ denote the anomalous magnetic moments of the decaying and emitted
baryons, respectively.

The tildes on $Q_6$ and $Q_7$ indicates that contributions of order
$(q/M_1)^2$ and higher have been subtracted. Also,
$Q_8,\ldots,Q_{25}$ have only contributions up to order $q/M_1$.

Although we have not made it explicit, in the above expressions
the primed form factors, containing the model dependence of
virtual RC should be used. This is valid to order $(\alpha/\pi)^2$
rearrangements. In the coefficients $Q_6,\ldots,Q_{15}$ we have
used the Harrington's form factors $F_i$, $G_i$. They are related
to the Dirac's form factors $f_i$, $g_i$ as $F_1 = f_1 + (1 +
M_2/M_1)f_2$, $G_1 = g_1 - (1 - M_2/M_1)g_2$, $F_2 = -2f_2$, $G_2
= -2g_2$, $F_3 = f_2 + f_3$, $G_3 = g_2 + g_3$.

\section{Collection of the $\Lambda_i$ functions}\label{appb}

Here we give the analytical expressions of the $\Lambda_i$
functions that appear in Sec.~III in the analytical form of the RC
to the polarized decay rate.
\begin{equation}
\Lambda_1 = -El \theta_0, \nonumber
\end{equation}
\begin{equation}
\Lambda_3 = \frac{E}{2} \left[ (\beta^2-1) \chi_{12} + 2 \chi_{11}
- \chi_{10} \right], \nonumber
\end{equation}
\begin{equation}
\Lambda_4 = \frac{El p_2^2}{2M_1}
\left[2Y_2-Y_3-\frac{2\theta_0}{E}\right], \nonumber
\end{equation}
\begin{equation}
\Lambda_5 = \frac{El}{2M_1}\left[p_2^2
Y_3+2Z_2+\frac{2p_2l^2}{E}Y_1\right], \nonumber
\end{equation}
\begin{equation}
\Lambda_6 = \frac{l}{M_1}\left[p_2^2\theta_0 + (E+E_\nu^0)Z_1 -
EZ_2 - l^2p_2Y_1\right], \nonumber
\end{equation}
\begin{equation}
\Lambda_7 = \frac{El}{2M_1}Z_1, \nonumber
\end{equation}
\begin{equation}
\Lambda_8 = -\frac{p_2^2l}{M_1} \theta_0, \nonumber
\end{equation}
\begin{equation}
\Lambda_9 = \frac{1}{2}l p_2^2\theta_3, \nonumber
\end{equation}
\begin{equation}
\Lambda_{10} = \frac{1}{2}l \zeta_{11}, \nonumber
\end{equation}
\begin{equation}
\Lambda_{11} = \frac{1}{2}\beta p_2^2(\gamma_0-E\theta_3),
\nonumber
\end{equation}
\begin{equation}
\Lambda_{12} = \frac{1}{2}\left[ -l \zeta_{10}-\beta
Z_3-\frac{X_3}{E}+\frac{1}{2}(\chi_{21}-\chi_{20}) +
\frac{E_\nu^0}{E} X_2 \right], \nonumber
\end{equation}
\begin{equation}
\Lambda_{13} =
\frac{lp_2^2}{2M_1}\left[Y_4-\frac{X_2}{El}-2\eta_0-
\frac{\chi_{21}}{2El} - \frac{E+E_\nu^0}{E} \left[
\frac{1}{2}\theta_7 + E(\theta_4-\theta_3) \right] \right],
\nonumber
\end{equation}
\begin{eqnarray}
\Lambda_{14} & = & -\frac{lp_2^2}{2M_1}\left[\gamma_0-\beta
l\theta_3+\frac{X_2}{El}-\eta_0\right]+ \frac12
\frac{E+E_\nu^0}{M_1E}X_3-\frac{X_4}{2M_1} \nonumber \\
&   & \mbox{} + \frac{p_2l}{M_1}
\left[-\frac{y_0}{E}X_2+\frac{\eta_0}{4}\left[l(y_0-1)-2p_2\right]
\right], \nonumber
\end{eqnarray}
\begin{eqnarray}
\Lambda_{15} & = & \frac{\beta(E+E_\nu^0)}{4M_1}
\left[p_2^2\theta_7+2p_2^2E(\theta_4-\theta_3)+2\zeta_{21}-\frac{2}{l}X_3
\right] + \frac{X_4}{2M_1}-\frac{p_2l^2}{4M_1}(y_0^2-1) \nonumber \\
&  & \mbox{} + \frac{p_2}{4M_1E}\left[4(ly_0+p_2)X_2+p_2\chi_{21}
\right], \nonumber
\end{eqnarray}
\begin{equation}
\Lambda_{16} = \frac{l}{4M_1}\zeta_{21}, \label{eq:e52} \nonumber
\end{equation}
\begin{equation}
\Lambda_{17} = -\frac{\beta p_2^2}{4M_1}\left[ \frac{\chi_{21}}{l}
-2E\eta_0 + (E+E_\nu^0) \left[\theta_7 + 2E(\theta_4-\theta_3)
\right] \right], \nonumber
\end{equation}
\begin{equation}
\Lambda_{18} = \frac{1}{4E}\left[X_3-2E_\nu^0 X_2\right],
\nonumber
\end{equation}
\begin{equation}
\Lambda_{19} = \frac{E}{M_1}\left[ \chi_{20}-\chi_{21}+2E_\nu^0
(\chi_{11}-\chi_{10}) +\beta (\zeta_{21}-2E_\nu^0 \zeta_{11})
\right], \nonumber
\end{equation}
\begin{eqnarray}
\Lambda_{20} & = & \frac{E}{M_1}\left[\frac{}{} \beta l^2p_2
(Y_5-Y_1) - (E_\nu^0+l \beta)(\chi_{11}-\chi_{10}- \beta
\zeta_{11}) \right. \nonumber \\
&  & \mbox{} + \left. \beta (E+E_\nu^0)(\zeta_{11}-\zeta_{10}) +
\frac12 l\beta p_2(1-y_0) (\theta_0+\eta_0) + \beta^2lp_2^2 I
\right], \nonumber
\end{eqnarray}
\begin{equation}
\Lambda_{21} = \frac{E}{M_1}\left[ \beta l^2p_2 (Y_5+Y_1) -
\frac{1}{2E}X_4 + lp_2^2Y_3 + l Z_2\right], \nonumber
\end{equation}
\begin{eqnarray}
\Lambda_{22} & = & \frac{E}{M_1}\left[ \frac{1}{2}\chi_{20}+\beta
(\beta E_\nu^0-l) \chi_{11} - \frac12(1+\beta^2)
\chi_{21}\right. \nonumber \\
&  & \mbox{} + \mbox{} \left. \beta
(E-E_\nu^0)(\zeta_{11}-\zeta_{10}) +\beta \zeta_{21}+\frac{1}{2}l
\beta p_2 (1-y_0^2) \right], \nonumber
\end{eqnarray}
\begin{eqnarray}
\Lambda_{23} & = & \frac{l}{M_1}\left[ \frac{X_4}{2l}+(\beta
E_\nu^0+l) \chi_{11}-(E+E_\nu^0) (\zeta_{11}-\zeta_{10}) +
\frac{1}{2}p_2l(1-y_0^2) \right], \nonumber
\end{eqnarray}
\begin{equation}
\Lambda_{24} = \frac{l}{M_1}\left[ \frac{1}{2}p_2l (y_0-1)
\theta_0+E_\nu^0 \zeta_{10}-(E_\nu^0+l \beta) \zeta_{11} - \beta
lp_2^2I \right], \nonumber
\end{equation}
\begin{eqnarray}
\Lambda_{25} & = & \frac{p_2l}{4M_1}\left[
2y_0\chi_{11}+l(\theta_0+2\eta_0)(1-y_0) +\frac{2}{p_2}
(E+E_\nu^0-\beta p_2y_0)
\zeta_{11}\right. \nonumber\\
&  & \mbox{} - \left. \frac{2E}{p_2}\zeta_{10} - (E_\nu^0+\beta l)
\frac{\chi_{21}}{p_2l} + \beta (E+E_\nu^0) \frac{\zeta_{21}}{p_2l}
+ N_1 + 2\beta lp_2I \right], \nonumber
\end{eqnarray}
\begin{equation}
\Lambda_{26} = \frac{1}{4M_1}\left[(E_\nu^0-l\beta)
\chi_{21}-\beta (E_\nu^0-E)\zeta_{21}+ p_2lN_1 - \chi_{31} +
\beta\zeta_{31} \right], \nonumber
\end{equation}
\begin{equation}
\Lambda_{27} = \frac{1}{4M_1}\left[ E_\nu^0\chi_{20} + p_2l N_2
\right], \nonumber
\end{equation}
and
\begin{equation}
\Lambda_{28} = - \frac{p_2l}{4M_1}N_2. \nonumber
\end{equation}

In these $\Lambda_i$ we introduced the definitions
\begin{equation}
X_2 = \frac{m^2}{E}\chi_{12}-E\chi_{11}-\frac{1}{2}\chi_{21},
\nonumber
\end{equation}
\begin{equation}
X_3 = \frac{m^2}{E}\chi_{22}-E\chi_{21}-\frac{1}{2}\chi_{31},
\nonumber
\end{equation}
\begin{equation}
X_4 = \frac{m^2}{E}\chi_{21}-E\chi_{20}, \nonumber
\end{equation}
\begin{equation}
Y_1 = \theta_{19} - \frac{l}{p_2}\theta_{20}-\frac{E_\nu^0}{p_2}
\theta_{10}, \nonumber
\end{equation}
\begin{equation}
Y_2 = 2\theta_3+(\beta^2-1) \theta_2-\theta_4, \nonumber
\end{equation}
\begin{equation}
Y_3 = (\beta^2-1) \theta_3+\theta_4+\beta \theta_5, \nonumber
\end{equation}
\begin{equation}
Y_4 = \gamma_0-3E\theta_3+E\theta_4+2\theta_7-3l\theta_5
+(1-\beta^2)(2E\theta_2-\theta_6) +\frac{1}{2E}\theta_9, \nonumber
\end{equation}
\begin{equation}
Y_5 = \theta_{19} - \frac{1}{2p_2}
\left[\theta_{21}+\frac{E_\nu^0}{l}\theta_{14}\right]
-2Y_1+y_0\theta_5, \nonumber
\end{equation}
\begin{equation}
Z_1 = (\beta^2-1) \zeta_{12}+2\zeta_{11}-\zeta_{10}, \nonumber
\end{equation}
\begin{equation}
Z_2 = (\beta^2-1) \zeta_{11}+\zeta_{10}, \nonumber
\end{equation}
\begin{equation}
Z_3 = \frac{m^2}{E}\zeta_{12}-E\zeta_{11}-\frac{1}{2}\zeta_{21},
\nonumber
\end{equation}
\begin{equation}
N_1 = l \eta_0 \left[\frac32 (1-y_0) - \frac{p_2}{l}\right],
\nonumber
\end{equation}
\begin{equation}
N_2 = l \eta_0 \left[\frac12 (y_0-1) + \frac{p_2}{l}\right],
\nonumber
\end{equation}
and
\begin{equation}
\gamma_0 = -\frac{m^2}{E} \theta_2 + E\theta_3 + \frac12\theta_7.
\nonumber
\end{equation}
$\eta_0$ is defined as $\eta_0 = 1+y_0$. All of the quantities
$\zeta_{pq}$, $\chi_{mn}$ except $\zeta_{31}$ as functions of the
$\theta_1,\ldots,\theta_{18}$ come from previous work
\cite{rfm97}. The $\theta_0,\ldots,\theta_{18}$ are found in
Refs.~\cite{rfm97,tun93}. $\zeta_{31}$, $I$, $\theta_{19},\ldots$,
$\theta_{22}$, are all new functions and they are given by
\begin{eqnarray}
I & = & \frac{3}{2\beta p_2}(E+E_\nu^0)(\theta_{13}-\theta_{12}) +
\frac12 y_0\theta_{12} + \frac{\beta E_\nu^0+l-p_2}
{2\beta p_2^2} \theta_0 + \frac{\eta_0 E_\nu^0}{p_2^2} \nonumber \\
&  & \mbox{} +
\frac{1}{2p_2^2\beta^2}\left[3(E_\nu^0)^2-l^2+3E(E+2E_\nu^0)\right](\theta_3-\theta_4-\beta
\theta_5)
+ \frac{EE_\nu^0}{p_2^2}(\theta_4-\theta_3) \nonumber \\
&  & \mbox{} - \frac{(E_\nu^0)^2}{2p_2^2}\theta_3 +
\frac{3E}{2p_2}Y_1 - \frac{3E(E+E_\nu^0)}{2p_2^2}\theta_{10} +
\frac{1}{2\beta^2}Y_3, \nonumber
\end{eqnarray}
\begin{eqnarray} \zeta_{31} & = & p_2ly_0 \left[ 2(3E^2-l^2)
\theta_3-6E^2(\theta_4+\beta \theta_5) + \theta_9 \right]
- 30lE^2p_2\theta_{13} - 30l^2Ep_2 \theta_{19} \nonumber \\
&  & \mbox{} -\frac{6l^3}{\beta^4} \left[ 5(l + \beta E_\nu^0) +
3\beta^2(p_2y_0-l) \right] (\theta_3-\theta_4-\beta
\theta_5) - 18l^2EE_\nu^0(\theta_4-\theta_3) \nonumber \\
&  & \mbox{} + 6p_2l^3y_0 \theta_3 + 30lE^2 (l + \beta E_\nu^0)
\theta_{10} + 30El^3 \theta_{20} - \frac12 \theta_{22}
\nonumber \\
&  & \mbox{} -6p_2\left[l^2E(\beta^2-5) - \frac{2lp_2^2+2\beta
p_2ly_0(E+E_\nu^0)}{b^+b^-}\right] \theta_{12}. \nonumber
\end{eqnarray}
The functions $\theta_{19},\ldots,\theta_{22}$ are
\begin{eqnarray}
\theta_{19} & = &  \frac{1}{p_2}(T_{19}^+ + T_{19}^-), \nonumber\\
\theta_{20} & = &  \frac{1}{p_2}(T_{20}^+ + T_{20}^-), \nonumber\\
\theta_{21} & = &  \frac{1}{p_2}(T_{21}^+ + T_{21}^-), \nonumber \\
\theta_{22} & = & \frac{1}{p_2} (T_{22}^+ + T_{22}^-), \nonumber
\end{eqnarray}
with
\begin{equation}
T_{19}^\pm = \frac{1}{3p_2} \left[p_2 - l + \frac12 E_\nu^0
(x_0^2-3) x_0 \right], \nonumber
\end{equation}
\begin{eqnarray}
T_{20}^\pm & = & \frac{1}{4} \left[ x_0^4 \ln \left| \frac{1\pm
x_0}{\pm x_0 \pm a^\pm} \right| + (a^\pm)^4 \ln
\left| \frac{\pm x_0 \pm a^\pm}{1\pm a^\pm} \right| + \ln \left| \frac{1\pm a^\pm}{1\pm x_0}\right| \right. \nonumber \\
&  & \mbox{} - \left. \frac{1}{3}(1\mp x_0^3) (1\mp a^\pm) +
\frac{1}{2}(1-x_0^2) \left[ 1- (a^\pm)^2 \right] - (1\mp x_0)
\left[ 1\mp (a^\pm)^3\right] \right], \nonumber
\end{eqnarray}
\begin{eqnarray}
T_{21}^\pm = \frac{2}{3} \left[p_2-l-E_\nu^0x_0^3 \right] \mp p_2y_0^\pm a^\pm \left[a^\pm I_2^\pm-2\right]
 - E_\nu^0x_0^\pm\left[2a^\pm x_0-x_0^2+1+(a^\pm)^2J_2^\pm \right], \nonumber
\end{eqnarray}
\begin{eqnarray}
\frac{T_{22}^\pm}{2l} & = &
\frac{2}{\beta}\left[(E_\nu^0)^3x_0+(l-p_2)^3\right]-6p_2E\eta_0(p_2+l)+12\eta_0p_2la^\pm
\left[
\pm E+p_2 \frac{y_0^\pm}{b^\pm} \right] \nonumber \\
&  & \mbox{} + \frac{[E_\nu^0(1-\beta x_0)]^3}{\beta (b^\pm)^2} J_1 + \frac{1}{b^\pm} \left[\mp 3(a^\pm \mp 1)\eta_0p_2l(\beta\eta_0p_2+2E+2E_\nu^0) \right] I_1\nonumber \\
&  & \mbox{} + \frac{1}{(b^\pm)^2}
\left[\mp(p_2\eta_0)^3\beta^2-\frac{1}{\beta}(E_\nu^0\beta+l-p_2)^3
 - 3p_2\eta_0(E_\nu^0\beta+l-p_2) (p_2\beta y_0+E_\nu^0+E)\right]I_1 \nonumber \\
&  & \mbox{} +
\left[\frac{3\eta_0p_2}{\beta}[E(l-p_2)+p_2(E_\nu^0\mp 2l \pm
2p_2)] -3 \frac{(a^\pm \mp 1)}{b^\pm}
p_2^2l[\pm \eta_0^2 + 2a^\pm (\eta_0 + y_0^\pm)] \right] I_2^\pm \nonumber \\
&  & \mbox{} + \frac{1}{(b^\pm)^2} \left[\mp \beta (p_2\eta_0)^3
-p_2^2(a^\pm \mp1)^2[3(E_\nu^0\beta+l-p_2)-2p_2\beta(a^\pm \mp
1)]\right] I_2^\pm
\nonumber \\
&  & \mbox{} + \frac{1}{(b^\pm)^2} \left[ -
\frac{3p_2\eta_0}{\beta}(E_\nu^0\beta+l-p_2)
(p_2\beta y_0+E_\nu^0+E) \right] I_2^\pm - \frac{\left(E_\nu^0 x_0^\pm\right)^3}{b^\pm}\left(J_3^\pm \pm I_3^\pm \right)\nonumber \\
&  & \mbox{} +
\left[\frac{(E_\nu^0)^3(x_0^\pm)^2}{(b^\pm)^2}(3-\beta x_0+2\beta
a^\pm)-6p_2lE_\nu^0\left(\frac{a^\pm y_0^\pm
x_0^\pm}{b^\pm}\right) \right] J_2^\pm, \nonumber
\end{eqnarray}
where $y_0^\pm = y_0\pm a^\pm$, $b^\pm=1+\beta a^\pm$, and
$x_0^\pm=x_0+a^\pm$. The functions $a^\pm$, $x_0$, $I_1$, $I_2^\pm$,
$I_3^\pm$, $J_1$, $J_2^\pm$, and $J_3^\pm$ are found in
Ref.~\cite{rfm97}.

\section{Collection of the $\rho_i$ functions}\label{appc}

Here we give the analytical results for the $\rho_i$ functions
after performing the integrals displayed in
Eqs.~(\ref{eq:e21})-(\ref{eq:e28}).
\begin{eqnarray}
\rho_I & = & \frac{l}{M_1}\left\{E_\nu^0 p_2(y_0-1)\left[
\frac{1-\beta^2}{2\beta}\theta_0-\beta \eta_0\right]
+ (E+E_\nu^0) (\zeta_{10}-\zeta_{11}) \right. \nonumber \\
&  & \mbox{} \left. + \frac{p_2l}{2}(y_0-1)\theta_0 + l^2p_2Y_1-
\beta l p_2^2I\right\}, \label{eq:e87} \nonumber
\end{eqnarray}
\begin{equation}
\rho_I^\prime = \frac{El}{M_1}\left[ (E_\nu^0+E)(\beta
p_2\theta_{12}-\theta_0) -p_2l\theta_{13}\right], \label{eq:e88}
\nonumber
\end{equation}
\begin{eqnarray}
\rho_{II} & = &
\frac{p_2l}{2M_1}\left[\frac{p_2E_\nu^0}{2}\theta_4+\frac{\eta_0}{4}\left[(y_0-1)
(2E_\nu^0\beta-3l)-2p_2
\right] -\frac{E_\nu^0}{2\beta p_2}\chi_{10}+\frac{1}{4\beta p_2}\chi_{20} \right. \nonumber \\
&  & \mbox{} +
\frac{1}{2}\left[y_0+\frac{E_\nu^0(E_\nu^0+E)}{p_2l}\right]\chi_{11}
- \frac{E_\nu^0+E}{4p_2l}\chi_{21}+\
\frac{y_0-1}{4\beta}\left[(\beta^2-1) E_\nu^0- \beta l \right]\theta_0 \nonumber \\
&  & \mbox{} + \left.
\frac{E_\nu^0-E}{2p_2}\zeta_{10}+\frac{1}{2}\left[\frac{E^2-(E_\nu^0)^2}{Ep_2}-\beta
y_0\right] \zeta_{11} +
\frac{E_\nu^0+E}{4Ep_2}\zeta_{21}+\frac{\beta p_2l}{2}I\right],
\label{eq:e89} \nonumber
\end{eqnarray}
\begin{eqnarray}
\rho_{II}^\prime & = & \frac{p_2l}{8M_1} \left[2(p_2+l
y_0)(\theta_0-\beta p_2\theta_{12}) + l(1-y_0^2) +2l^2 Y_5\right.
\nonumber \\
&  & \mbox{} + \left. \frac{E+E_\nu^0}{p_2l} \left[\beta
\zeta_{21}+\chi_{20} - \chi_{21}\right] \right], \label{eq:e90}
\nonumber
\end{eqnarray}
\begin{eqnarray}
\rho_{III} & = &\frac{p_2l}{2M_1} \left[\frac{E}{2p_2l}\left[
2E_\nu^0(1-\beta^2) \chi_{11}-2E_\nu^0\chi_{10}-(1-\beta^2)
\chi_{21}+\chi_{20}\right] \right. \nonumber \\
&  & \mbox{} - \left.
\frac{2E_\nu^0}{p_2}\zeta_{10}+\frac{l(y_0^2-1)}{2}+\frac{E_\nu^0+\beta
l}{p_2}\zeta_{11} + \beta p_2
lI - \frac{l(y_0-1)}{2}\theta_0 \right] \nonumber \\
&  & \mbox{} - \frac{p_2l}{4M_1} \eta_0\left[ l (y_0-1) +
2p_2\right] - \frac{E_\nu^0}{2M_1} \chi_{20}, \label{eq:e91}
\nonumber
\end{eqnarray}
and
\begin{eqnarray}
\rho_{III}^\prime & = & \frac{p_2l}{2M_1} \left[E(p_2-l y_0)
\theta_4 + \frac{EE_\nu^0}{p_2}\eta_0+l(p_2+l y_0) \theta_5
-\frac{p_2m^2}{E}\theta_3 - \frac{E}{p_2}\left[2-\beta^2+\frac{E_\nu^0}{E}\right] \zeta_{11}\right. \nonumber \\
&  & \mbox{} +\frac{3E}{p_2}\zeta_{10} +
\frac{E}{p_2l}(E_\nu^0+\beta l) \chi_{11}- \frac{1-\beta^2}{2\beta
p_2}\chi_{21}
+\frac{E}{p_2l}\chi_{20}-\frac{EE_\nu^0}{p_2l}\chi_{10} \nonumber \\
&  & \mbox{} -\left. \frac{l^2}{2p_2} (\theta_{21}-2l\theta_{20})
- \frac{l}{2p_2}(E_\nu^0-E)(\theta_{14}-2l \theta_{10}) +
\frac{\eta_0}{2} \left[ l (y_0-1) + 2p_2\right] \right]. \nonumber
\label{eq:e92}
\end{eqnarray}
All the algebraic expressions that appear in these $\rho_i$ and
$\rho_i^\prime$ are defined in Appendix \ref{appb}.

\section{New and previous analytical integrals}\label{appd}

In this Appendix, we give a brief discussion that allows the
identification of previous and new analytical integrals over the
photon variables that emerge in the present calculation. Such
integrals can all be put into the general form
\begin{equation}
\int_{-1}^{y_0} dy F^p \int_{-1}^1 dx \int_0^{2\pi} d\varphi_k
\frac{x^r {\hat {\mathbf s}_1} \cdot {\hat {\mathbf a}}}{D^m
(1-\beta x)^n}, \label{eq:d3} \nonumber
\end{equation}
with ${\hat {\mathbf a}}={\hat {\mathbf p}_2}$, ${\hat {\mathbf
l}}$, ${\hat {\mathbf k}}$ and $x={\hat {\mathbf l}}\cdot
{\hat{\mathbf k}}$. $F$ and $D$ were defined after
Eq.~(\ref{eq:e16}). The powers of $x$, $F$, $D$, and $(1-\beta x)$
are denoted by $\{rpmn\}$. Since each one of these quantities is a
rotational scalar, one may change the orientation of the space
axes. For ${\hat {\mathbf s}}_1\cdot {\hat {\mathbf a}}$ we use
the rule \cite{rfm97}
\begin{equation}
{\hat {\mathbf s}_1} \cdot \hat {\mathbf a} \rightarrow ({\hat
{\mathbf s}_1} \cdot {\hat {\mathbf p}_2}) (\hat{\mathbf a} \cdot
{\hat {\mathbf p}_2}), \label{eq:b32} \nonumber
\end{equation}
where ${\hat {\mathbf a}}={\hat {\mathbf p}_2}$, ${\hat {\mathbf
l}}$, or ${\hat {\mathbf k}}$. This considerably simplifies the
calculation of such integrals. One can classify these integrals
into three groups. In the first one we can directly identify
integrals previously performed. This occurs for ${\hat {\mathbf
a}}={\hat {\mathbf p}}_2$ and $\{rpmn\}=\{0012,0011,0010\}$, then
one identifies $\theta_2$, $\theta_3$ and $\theta_4$ of
Ref.~\cite{rfm97}. For ${\hat {\mathbf a}}={\hat {\mathbf l}}$ and
$\{rpmn\}=\{0010,0011,0012,0121\}$ one identifies the functions
$\zeta_{pq}$. For ${\hat {\mathbf a}}={\hat {\mathbf k}}$ and
$\{rpmn\}=\{0010,0120,0011,0012,0121,0122,0231\}$ one identifies
the functions $\chi_{pq}$.

In the second group are new integrals that can be expressed as
combinations of previous results in terms of $\eta_0=1+y_0$ and
$\theta_0$, $\theta_2,\ldots,\theta_{18}$ also of
Ref.~\cite{rfm97}. This occurs for ${\hat {\mathbf a}}={\hat
{\mathbf p}}_2$ and
$\{rpmn\}=\{0000,0001,00(-1)1,0111,0121,0122,0231,1010\}$, for
${\hat {\mathbf a}}={\hat {\mathbf l}}$ and
$\{rpmn\}=\{0000,0001,0002,0120,0101\}$, and for ${\hat {\mathbf
a}}={\hat {\mathbf k}}$ and $\{rpmn\}=\{0001,1010,0111,
0112,0230\}$. Omitting details, such combinations are accommodated
into the $\Lambda_i$ of Appendix \ref{appb}.

The third group contains, after applying the above rule, only four
new integrals with ${\hat {\mathbf a}}={\hat {\mathbf p}}_2$ and
the powers are $\{rpmn\}=\{0211,2120,1110,0331\}$. The first three
of them are straightforward. Explicitly they are
\begin{equation}
\int_{-1}^1 dx \frac{1}{1-\beta x} \int_{-1}^{y_0} dy
F^2\int_0^{2\pi} d\varphi_k \frac{1}{D} = 2\pi (2p_2l)^2 \left[I +
y_0^2\theta_3-\frac{2y_0}{p_2l}\zeta_{11} \right], \nonumber
\end{equation}
where
\begin{eqnarray}
I & = & \int_{-1}^1 dx \frac{1}{1-\beta x} \int_{-1}^{y_0} dy
\frac{y^2}{\sqrt R}, \nonumber
\end{eqnarray}
\begin{equation}
\int_{-1}^1 dx x^2 \int_{-1}^{y_0} dy F \int_0^{2\pi} d\varphi_k
\frac{1}{D^2} = 2\pi(\theta_{21}- 2l\theta_{20}),\nonumber
\end{equation}
and
\begin{equation}
\int_{-1}^1 dx x \int_{-1}^{y_0} dy F \int_0^{2\pi} d\varphi_k
\frac{1}{D} = 2\pi(2p_2l)(y_0\theta_5-Y_1). \nonumber
\end{equation}
$\theta_{20}$ and $\theta_{21}$ are new and they are listed below.
$\theta_5$ is found in \cite{rfm97} and $Y_1$ is found in Appendix
\ref{appb}. The computation of the fourth integral,
\begin{equation}
{\mathcal J} = \int_{-1}^1 dx \frac{1}{1-\beta x} \int_{-1}^{y_0}
dy F^3 \int_0^{2\pi} d\varphi_k \frac{1}{D^3}. \label{eq:j10}
\nonumber
\end{equation}
although long and tedious, can be performed by using standard techniques. The final result can be organized as
\begin{eqnarray}
\frac{{\mathcal J}}{(2\pi)(12l^3)} & = & \frac{1}{\beta^4}[5(l+\beta E_\nu^0)+3\beta^2(p_2y_0-l)](\theta_3-\theta_4
- \beta\theta_5) + \frac{3E_\nu^0}{\beta}(\theta_4-\theta_3) \nonumber \\
&  & \mbox{} -p_2y_0\theta_3-\frac{5}{\beta^2}(l+\beta E_\nu^0)\theta_{10} -5E\theta_{20} + \frac{1}{12l^3} \theta_{22}
+ \frac{5p_2}{\beta^2}\theta_{13}+\frac{5p_2}{\beta}\theta_{19}\nonumber \\
&  & \mbox{} + p_2\left[\frac{\beta^2-5}{\beta} -
\frac{2p_2^2+2\beta p_2y_0(E+E_\nu^0)}{l^2b^+b^-} \right]
\theta_{12}. \nonumber
\end{eqnarray}

The functions $\theta_{19},\ldots,\theta_{22}$ in the four new
integrals are
\begin{eqnarray*}
\theta_{19} & = & \int_{-1}^1 x \xi_4(x) dx , \\
\theta_{20} & = & \int_{-1}^1 x^3 \xi_1(x) dx , \\
\theta_{21} & = & \int_{-1}^1 x^2 \xi_2(x) dx ,
\end{eqnarray*}
and
\begin{equation}
\theta_{22} = \int_{-1}^1 \frac{\xi_6(x)}{1-\beta x} dx .
\nonumber
\end{equation}
the other $\xi_1(x)$, $\xi_2(x)$, and $\xi_4(x)$ are used in the
Ref.~\cite{tun91}, the function $\xi_6(x)$ is new and it reads
\begin{eqnarray}
\frac{\xi_6(x)}{2l} & = &
p_2^2\eta_0^3\left[\frac{1}{(x+a^-)^2}-\frac{1}{(x+a^+)^2}\right]
+ 3l\eta_0\left[
p_2\eta_0+2x(E_\nu^0+lx)\right]\left[\frac{a^-+1}{x+a^-}-\frac{a^+-1}{x+a^+}\right] \nonumber \\
&  & \mbox{} -
3\eta_0\left[p_2y_0+x(E_\nu^0+lx)\right]\left[E_\nu^0+(l-p_2)x\right]\left[\frac{1}{(x+a^-)^2}+
\frac{1}{(x+a^+)^2}\right] \nonumber \\
&  & \mbox{} + \frac{(E_\nu^0)^3}{p_2}
\left[|x-x_0|^3-(1+xd)^3\right]\left[\frac{1}{(x+a^-)^2}+\frac{1}{(x+a^+)^2}\right]
\nonumber \\
&  & \mbox{} -
6lp_2^2\left[-\frac{a^+y_0^+}{b^+(x+a^+)}-\frac{a^-y_0^-}{b^-(x+a^-)}\right](1-\beta
x) \xi_4(x), \nonumber
\end{eqnarray}
where $d=(l-p_2)/E_\nu^0$.

\end{document}